# Machine Learning based Radio Environment Map Estimation for Indoor Visible Light Communication

**Helena Serpi, Christina (Tanya) Politi**
Department of Informatics and Telecommunications, University of Peloponnese, Tripoli, Greece
Department of Electrical and Computer Engineering, University of Peloponnese, Patras, Greece
Correspondence: Helena Serpi, e.serpi@go.uop.gr

## Abstract

This paper presents a machine learning-based methodology for constructing optical Radio Environment Maps (REM) for indoor visible light communication (VLC) systems, where at first the REM estimation is evaluated on configurable simulated datasets generated by a physics-based indoor VLC simulator and then the selected method for estimating REM is benchmarked against alternative learning and interpolation-based methods, on experimental measurements. Thus, the evaluation of Decision Trees and Multi-Layer Perceptron (MLP) models on received signal strength (RSS) datasets, generated from different room geometries and transmitter configurations, shows that MLPs achieve superior accuracy-performance trade-offs through appropriate tuning of the training sample size, number of epochs and batch size. Then, experimental benchmarking, spatial validation, and robustness analysis demonstrate the predictive accuracy and generalization ability of MLP across the spatial domain. To address sparse measurement scenarios, synthetic data were generated using SMOGN (Synthetic Minority Over-Sampling Technique for Regression with Gaussian Noise) and Gaussian Process Regression (GPR), and their exploration showed that GPR-based augmentation can reduce the required number of real measurements by up to 83% while maintaining prediction accuracy. These results support the practical viability of data-driven REMs for indoor VLC applications such as network planning, localization, and digital twins.

## 1 Introduction

The transition toward the full realization of 6G makes the need for reliable, high-speed, and ultra-low-latency communication more critical than ever. Internet of Things applications, virtual reality, and autonomous systems [1, 2] have nearly tripled wireless traffic volumes compared to 2021. New network applications and services such as integrated sensing and communication (ISAC) [3] and digital twins [4] require seamless connectivity across land, air and sea, thus expanding the boundaries of existing communication paradigms. The way we perceive and use mobile networks must change radically since 6G is not a simple upgrade of previous generation but a fundamental transformation [5].

Optical Wireless Communication (OWC) refers to any wireless system that uses the optical spectrum, which is the UV, visible and infrared bands, and uses light as an information carrier. The development of OWC applications has a significant impact on agriculture, healthcare, culture, security, climate, energy, transportation and digital industries [6]. Hybrid visible light and infrared optical wireless network IoT solutions demonstrate the flexibility and scalability of OWC systems in smart environments [7]. Their EMI immunity makes them ideal for indoor applications where the use of traditional radio frequency (RF) communications is prohibited [8]. Visible light as part of hybrid RF-VLC communication systems complements and reduces the burden of (RF) communications [9, 10]. Realization of 6G demands a functioning universal architecture that can integrate all wireless access technologies and OWC will have a central role in it [11].

VLC is a subset of OWC that primarily involves one-way communication using visible light emitting diodes (LEDs), which serve as the light source in the transmitter. Although LEDs have a narrower bandwidth and less focused light beams compared to laser diodes, their low cost, eye safety, and ease of use make them ideal especially in indoor deployments where the light used for communication can also serve as a source of illumination [12]. Indoor VLC systems have gained attention for their potential to support applications such as positioning [13, 14] and human sensing [15], capabilities compatible with key pillars of 6G that are wireless link at all spectra, Artificial Intelligence and Internet of Everything [1,16].

In recent years, there has been a great boost of research activity on indoor VLC channel modeling [17-20] and Radio Environment Maps (REMs) have emerged as one way to bridge theoretical channel models with the practical decisions required for network deployment. Building such maps is not trivial. Measurements of quantities such as received signal strength (RSS) are highly dependent on surface reflectivity, occlusions, and the particular geometry of the room. Romero and Kim [21] emphasized the value of data-driven radio map estimation for spectrum cartography while recent research trends at radio environment mapping techniques, are based on encoder-decoder architecture [22] using U-Net convolutional neural networks [23] to estimate the propagation loss, or by extending an autoencoder with transfer learning [24] to introduce scalable path loss map predictions for wireless networks.

In earlier work [25], we investigated whether Decision Trees could approximate indoor VLC systems' behavior and generate their virtual representation. Present study not only introduces a neural-network-based alternative for REM construction, using a Multi-Layer Perceptron (MLP) architecture, but proposes a new methodology for data-driven REM construction where first learning-based models using configurable simulated datasets are evaluated and subsequently the selected approach is benchmarked against alternative learning-based and interpolation methods using experimental measurements. Synthetic data production is also proposed, as part of this study, when only few experimental measurements are available, and results remain of high accuracy and speed.

The remainder of this paper is organized as follows. Section 2 presents the indoor VLC system model and the construction of the corresponding power map. Section 3 briefly describes the machine learning algorithms and outlines the methodology and evaluation metric used for their application to VLC systems. Section 4 presents and discusses results obtained from simulation-generated data, while Section 5 evaluates the proposed MLP model using real experimental measurements and compares it against several baseline algorithms. Section 6 investigates the generation and use of synthetic data for REM construction. Finally, Section 7 summarizes the results and concludes the paper.

# 2 System Model

In this section, the indoor VLC system model is presented, detailing the transceivers' setup within the indoor environment and the channel propagation model used to create the optical power map of the system. This model is run to achieve a large set of values of the optical map and exact receiver/transmitter locations. These values are used for training our models and then as references to check the accuracy of predictions.

## 2.1 System setup

In this study, we choose a simplified model to represent indoor VLC system, because the primary purpose is not to approach realistic indoor VLC systems but to evaluate the methodology we introduce for data-driven Radio Environment Mapping (REM), under controlled and repeatable conditions to separately investigate the impact of data representation, sampling and learning strategies. Therefore, the effect of shadowing, dynamic blocking, and complex multipath propagation are explicitly excluded in the present study. In this context, VLC systems considered consist of either one or four LED transmitters placed on the ceiling of empty rooms and a single photodetector (PD) receiver. LED transmitters are oriented downward and serve both illumination and communication purposes, while the PD is receiving data and faces upward, with its normal axis perpendicular to the floor, as illustrated in Figure 1. For the optical wireless channel, both line-of-sight (LoS) transmission and single-bounce wall reflections contributing to non-line-of-sight (NLoS) propagation are considered.

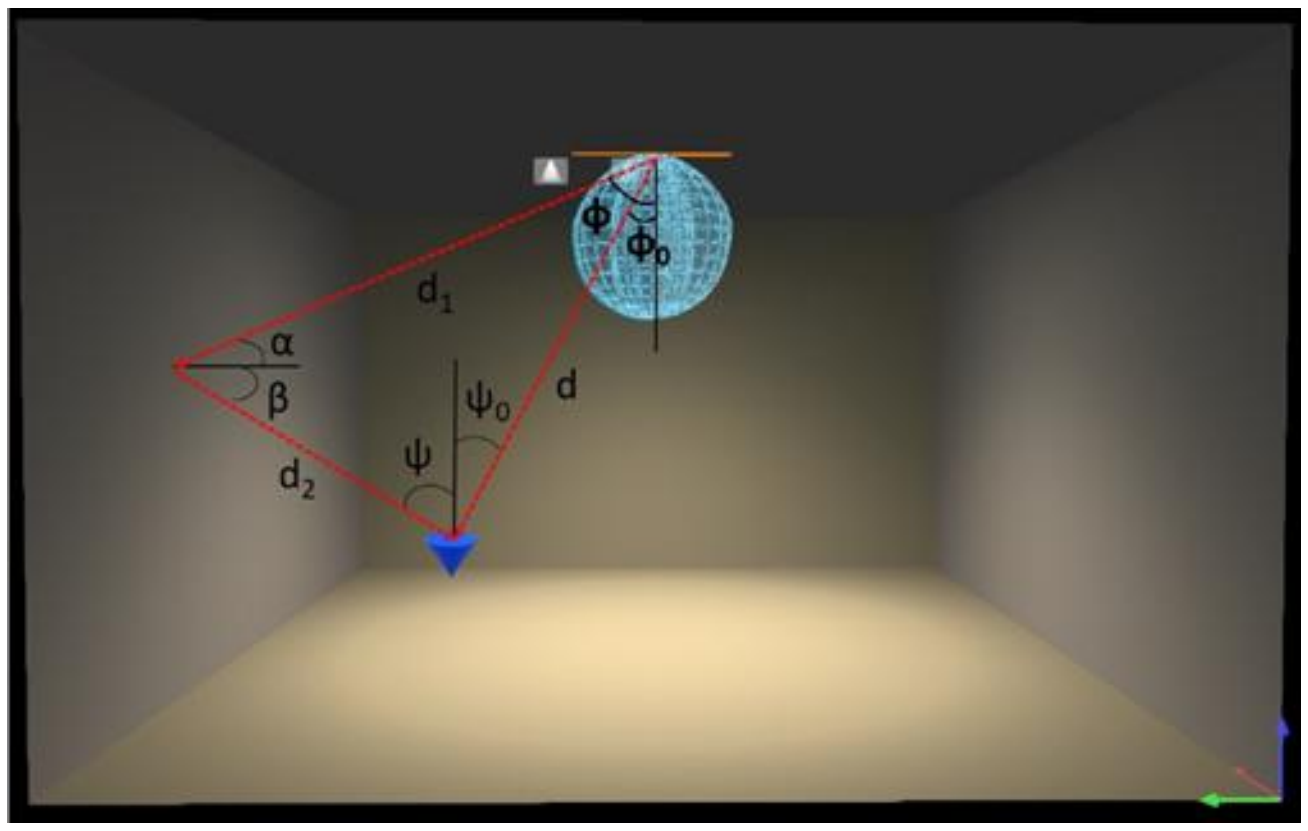

**FIGURE 1** VLC system setup with a single LED mounted on the center of the ceiling with an indicative Lambertian radiation lobe of m=1. Image produced with DIALux evo [26].

## 2.2 Indoor VLC channel propagation model

Assuming the optical channel is distortionless and static, i.e., it has the same gain for all frequencies of interest, the optical power PR arriving at the PD can be expressed as [27]

$$P_R = P_T\, H_{LOS}(0) + \int P_T\, dH_{ref}(0) \qquad (1)$$

where PT is the emitted optical power by LEDs, $H_{LOS}(0)$ and $H_{ref}(0)$ the zero-frequency (DC) value of the direct and reflected paths.

The zero frequency responses can be expressed as

$$H(0) = \int_{-\infty}^{\infty} h(t)dt \qquad (2)$$

where *h(t)* defined as the baseband Channel Impulse Response. [28]

Path Loss of unobstructed diffuse configurations can be estimated as [29]

$$PL_{Diffuse} = -10log_{10}\left(\int_{-\infty}^{\infty} h(t)dt\right) \qquad (3)$$

LEDs, PDs and diffuse reflecting surfaces are simulated using a generalized Lambertian model.

For the LOS path

$$H(0) = \frac{(m+1)A_{PD}}{2\pi d^2} cos^m(\varphi_0)\, T_s(\psi_0)\, g(\psi_0)\, cos(\psi_0) \quad (4)$$

when $0 \leq \psi_0 \leq \psi_c$ and 0 when $\psi_0 > \psi_c$, where Lambertian order *m* that expresses the directionality parameter of LED Tx is defined by

$$m = \frac{-\ln 2}{\ln cos(\Phi_{1/2})} = \frac{-\log_{10} 2}{\log_{10} cos(\Phi_{1/2})} \qquad (5)$$

Half-power semi-angle of Tx is $\Phi_{1/2}$, $A_{PD}$ and $\psi_c$ are the active area and field of view (FOV) semi-angle of PD, *d* is the distance from LED to the PD location, $T_s(\psi 0)$ and $g(\psi 0)$ are the optical filter gain and the optical concentrator gain, respectively.

For Single bounce reflection NLOS

$$dH_{ref}(0) = \frac{(m+1)A_{PD}}{2\pi d_1^2 d_2^2}\, \rho\, dA_{wall} cos^m(\varphi)\, cos(\alpha) \cdot cos(\beta)\, T_s(\psi)\, g(\psi)\, cos(\psi) \qquad (6)$$

when $0 \leq \psi \leq \psi_c$ and 0 when $\psi > \psi_c$, $\alpha$ and $\beta$ are the angles of incidence and irradiance to and from a reflective area on the wall, $d_1$ and $d_2$ are the distances between Tx and the wall, and the wall and PD, respectively (Figure 1). $dA_{wall}$ is the size of the wall's reflective area and $\rho$ is the reflectance factor.

For multi bounce reflections NLOS [27] of 3, 4, or more bounces with nested loops cause computation time to grow exponentially. To optimize this, we use the Matrix-based Radiosity Method [30, 31]. Here, instead of tracking individual rays, we divide the room's walls, floor, and ceiling into a grid of "reflecting patches." We then compute the channel gain between these patches using large, highly optimized matrices. This turns an impossible looping task into a lightning-fast matrix multiplication that can handle any number of bounces almost instantly.

Total power detected from the PD is the linear summation of the power contributions from each individual LED (if more than one) [32]

$$P_R = \sum_{i=1}^{N} P_{LOS,i} + \sum_{i=1}^{N} P_{NLOS,i} + n_G \quad (7)$$

where $P_{LOS,i}$ and $P_{NLOS,i}$ are the powers that arrive to PD via LoS and NLoS paths from the i[th] Tx, respectively. Specifically, $P_{NLOS,i}$ represents the total accumulated power from all modeled reflections (e.g., the sum of the 1st through the 10th bounce) originating from that i[th] Tx. The total noise $n_G$ that originates from thermal noise, dark current, and shot noise induced by the signal and background radiation is represented as zero-mean additive white Gaussian noise with a power level (variance) of $\sigma^2$.

Using this optimized VLC room simulator, we calculate the received optical power by the Rx that is located on the diagonal of the plane 1m above the floor in a room 5m × 5m × 3m for VLC systems of 1, 3 and 4 LEDs for NLOS paths of one and 10 bounces. All selected LEDs are mounted on the ceiling, and specifically single LED is mounted on the center of the ceiling, 4 LEDs are placed symmetrically, while 3 LEDs have the same placement configuration as the 4 LEDs system but with one corner completely dark. Notice the steep signal drop near the end of the diagonal due to the deactivated LED in plot of Figure 2c. As shown in the plots of Figure 2, overall power is enhanced where more bounces are considered.

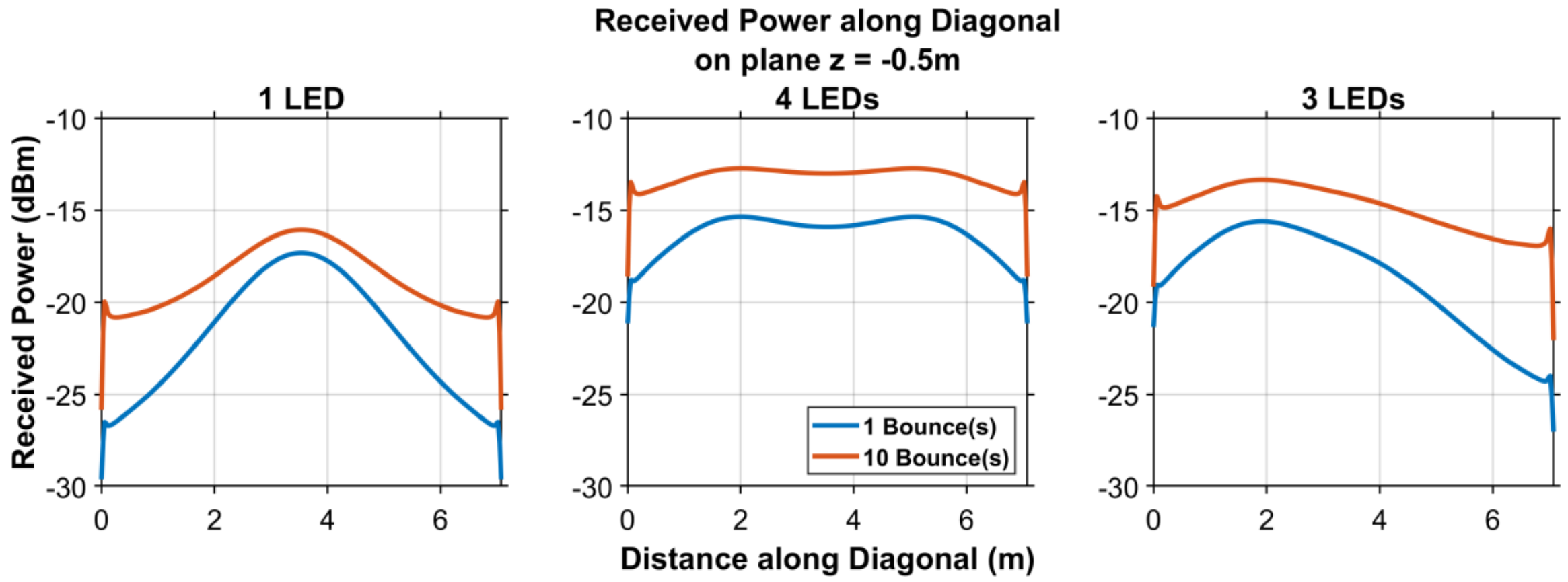

**FIGURE 2** VLC system setup in a 5m × 5m × 3m room with 1, 4 and 3 LEDs and the received optical power by the Rx that is located on the diagonal of the plane 1m above the floor for NLOS paths with 1 and 10 bounces.

## 2.3 Optical Radio Environment Maps

Spectrum Cartography, also known as REM, is a valuable tool to understand the radio channel characteristics of a geographical area or an indoor location, especially in environments like industrial sites and hospitals, where it is essential to function ultra-reliable communication systems [33]. REMs present measurements of various key quantities describing radio communication environments, like RSS, signal-to-noise ratio (SNR), channel attenuation, at different points within a geographical area. REM for a specific location can be created either using direct field measurements for the selected quantity or using the quantity values determined by Physics-based models. [34]

The signal received at any given point in the space under consideration depends on the signal being transmitted and the behavior of the propagation medium between the transmitter and receiver. Signal strength maps present the strength of the received signal, essentially recording the effect of the channel on transmissions from all active sources. One advantage of these maps is that they can be constructed without needing to know the number, positions, or power levels of the transmitters, making them particularly useful in situations with many mobile transmitters, such as in D2D communications or cellular uplinks. Signal strength maps can vary in the level of detail they provide, resulting in different types such as coverage, outage probability, and power maps. Among these, power maps offer more detailed information, making them valuable for tasks like network planning, trajectory optimization, transmitter localization, and fingerprint-based localization. [21]

In indoor VLC systems, communication channel characteristics mainly depend on room geometry and transmitter/receiver relative positions. In this paper we use the multipath propagation model that is reflected in (4) to (7) to calculate the received power at every spatial location and construct the optical power map of the system.

## 3. Indoor VLC System Prototyping using Machine Learning Methods

In our previous work [25] we used bagging, boosting and stacking on randomized decision trees, to overcome the high computation time needed to calculate the RSS value at each 3D location inside a room with a VLC system. In this study ANNs are used to construct optical power maps and extensively compare Decision Trees and MLP models aiming for robust, accurate and efficient data-driven modelling of indoor VLC systems. In the following subsections, the implemented methodology is presented, a brief description of Machine Learning algorithms and metrics used to evaluate the performance of the ML models

### 3.1 Methodology

Various Machine Learning methods, like Extra-Trees, or AdaBoost with Extra Trees Regressor estimator and MLP with two hidden layers are deployed for modelling the indoor VLC systems. The optical power received in a set of locations is the input data to the chosen ML methods.

Hyper-parameters, that are set before training of an ML model, are the variables that control its behavior [35]. It is common practice to tune the hyper-parameters of a model by searching for the best cross-validation score in the hyper-parameter space. Hyperparameters for Decision Trees include factors such as how deep the tree is allowed to grow, the minimum sample count necessary for node splitting, and the smallest permissible sample size for terminal nodes. For ANNs, structural hyperparameters include the quantity of hidden layers and their respective neuron counts, or variables that specify how the network is trained like learning rate, batch size, dropout rate and epochs are some of Neural Networks' hyper-parameters.

To achieve models with high prediction accuracy and reduced training and prediction time, we examine how varying certain hyperparameters impacts the performance of the selected models, allowing us to compare their accuracy and efficiency.

## 3.2 Machine Learning methods description

In statistics, system modeling is the representation of existing relationships between the properties of one system. ML helps to build system models through regression analysis that reveals relationships between system variables [36]. Once such a relationship is determined, the model can be used to predict future values of specific system parameters like RSS, SINR, and channel gain.

### 3.2.1 Decision Trees (DTs)

Decision Trees constitute a flexible type of supervised learning method commonly applied to classification and regression problems [37]. They create a series of simple decision rules based on the input features and based on these rules they make prediction of the target variable. A disadvantage of decision trees is that they are highly sensitive to small changes in the data and their tendency to overreact to them can make their predictions unstable. For this reason, they are often used in combination with other trees as part of ensemble methods that help reduce variance and improve reliability. Ensemble techniques combine many individual models, which in literature are often referred to as weak learners. During their training, they retain the strengths of each model, resulting in improved prediction accuracy [38]. The three most widely used ensemble strategies are bootstrap aggregating (bagging), boosting, and stacking. Bagging technique employs statistical resampling to train an ensemble of models on diverse data subsets, subsequently combining their predictions to enhance overall model stability and accuracy. Well-known examples include Random Forests and Extra Trees, which both rely on building numerous randomized decision trees [38]. Boosting takes a different approach by building models one after another. Each new model 'learns from the mistakes' of the one before it, focusing its efforts on the most difficult data points. A key algorithm here is AdaBoost (Adaptive Boosting) [39], which assigns more weight to the misclassified samples in each round and combines the outputs of all models using a weighted approach to make the final prediction. Finally, Stacking trains a second-level model, often called meta-learner, on the predictions of several base models [40].

### 3.2.2 Artificial Neural Networks (ANNs)

For problems of complex and nonlinear mappings and large datasets available the use of ANNs is the appropriate solution that outperforms other ML algorithms [41]. The great potential presented by ANNs and especially Deep Neural Networks (DNNs) in regression analysis problems has been increasingly recognized in recent years. Specifically for next generation network physical layer and channel models, these methods can deal in real time with complex and unknown channel models. Since the channels are too complex to be accurately described by fixed mathematical models, there is a need for algorithms capable of performing communication tasks without relying on predefined channel assumptions. The "learned" algorithms of ANNs are perfect candidates for modelling physical layer communications [42].

The ANN imitates the human brain function. Neuron is the atomic unit of a neural network, which is an interconnected network of neurons. Each neuron performs a simple mathematical operation that is defined by the chosen activation function, which endows ANNs with their characteristic ability to approximate non-linear complex functional mappings between inputs and outputs. It converts the weighted sum of input signals of a neuron into the output signal. The basic architecture for ANNs necessitates three distinct layer types: an input interface, computational hidden layers, and an output transformation layer [43]. MLP is a fully connected feedforward neural network meaning that neural signals propagate exclusively forward, i.e. from input nodes to output nodes through hidden layers and that all neurons in one layer are projected to every neuron in the next layer.

Important ANN terminology is as follows:

- Sample: a single row of the input dataset
- Epoch: a full cycle of training on all available samples
- Batch: number of samples processed together in one forward pass
- Learning rate: a hyperparameter that controls the step size of weight updates during training, i.e. how much to adjust the network weights in response to the calculated gradient
- Loss Function: measures the error for a single training example and quantifies how far the network's prediction is from the actual target for one sample
- Weights / Bias: trainable parameters within the network that govern signal transformation between neural units. These values are iteratively optimized during backpropagation to minimize prediction error.

Frank Rosenblatt first introduced the concept of "backpropagation error correction" in 1962, but its full implementation, which is the modern approach, was done by D. E. Rumelhart et al [44]. The method incorporates gradient descent and its variants like stochastic gradient descent. Backpropagation is essential in training ANNs and relies on the selected Loss function and the Gradient of the loss function. The direction and the steepness of loss function's slope help to adjust network parameters properly to get nearer to the global minimum and minimize the prediction error.

Optimizers in neural networks affect the accuracy and training speed of the model as their algorithms guide the process of discovering the optimal set of weights and learning rates that drive the loss toward its minimum [45]. The most used optimizers include: Gradient Descent, that is the simplest optimization algorithm, modifies the weights by progressing along the negative gradient of the loss, gradually reducing the error. Stochastic Gradient Descent (SGD) on the other hand, updates model parameters by computing the gradient using a randomly (hence stochastic) selected subset of the data rather than the entire dataset, allowing more frequent updates and often faster convergence. Adagrad changes the learning rate for each parameter individually, depending on the historical magnitude of its slopes. Root Mean Square Propagation (RMSProp) dynamically changes the learning rate for each parameter using a moving average of recent squared slope values. It divides the learning rate by the square root of this moving average, which helps stabilize and speed up the optimization process. The Adam (Adaptive Moment Estimation) optimizer is a powerful and widely used optimization method that combines the advantages of AdaGrad and RMSProp. It calculates moving averages of both the slopes and their squared values and dynamically adapts the learning rate for each parameter throughout the training.

### 3.3 Metrics

Metric for comparison of the ML models' performance is Mean Absolute Error (MAE) that computes the arithmetic mean of unsigned prediction deviations from ground truth values throughout the samples. Average training time required for the models to converge, and average inference time are as well calculated.

## 4 MLP based REM with Simulation Data

The ML algorithms are implemented using scikit-learn and Tensorflow, while data analysis and visualization are carried out using Python libraries like pandas and matplotlib, all running on Python version 3.12.

### 4.1 Datasets generation and Simulation parameters

Table 1 lists the values of key parameters used for modelling VLC systems, like those used in literature [Table8.4 in [27], 46].

**TABLE 1** Simulation Parameters.

| Parameter | Value |
|---|---|
| Room sizes | $3 \times 3 \times 2.8$ m$^3$<br>$5 \times 5 \times 3$ m$^3$<br>$6.5 \times 6.5 \times 3.5$ m$^3$<br>(3-7) × (3-7) × 3 m$^3$ |
| Location of LEDs (Txs) | center of the ceiling for Single LED<br>in the center of each quarter of the ceiling for Four LEDs |
| Area of PhotoDetector (PD) | 10-4 m$^2$ |
| Half Power Angle of Tx (HPA) | 60$^o$ |
| Responsitivity of PD | 1.0 |
| Rx's Field of view (FOV) | 85$^o$ |
| Transmitted power | 1000mW |
| Gain of optical filter | 1.0 |
| Refractive index of lens at the PD | 1.5 |
| Reflection factor of walls | 0.8 |

Large sets of RSS values at random 3D positions inside the rooms under investigation were estimated using the model presented in section 2.2 [27]. Datasets of 125,000 samples (50 random locations per dimension) were created for each VLC system, namely one or four LED transmitters in small room (3×3×2.8 m$^3$), in mid-sized room (5×5×3 m$^3$) and in big room (6.5×6.5×3.5 m$^3$). We generated also, datasets of 400,000 samples (20 random locations per x, y dimension, 10 random per z dimension, and 10 per room length and width) for rooms with 3m height and varying length and width between 3 and 7m. The PD is randomly placed at heights ranging from floor level (0 m) up to 1.7 m, reflecting realistic positioning in typical indoor environments. The input dataset contains four columns: the first three correspond to the features, namely the receiver's coordinates in the x, y, and z dimensions, while the last column holds the target RSS values in dBm, computed using equation (7). In the case of rooms with variable length and width two more features are added where the length lx and the width ly of the room are stored. Finally, datasets of “real values” were created for prediction from the ML models, that consist of 500 randomly selected samples for fixed size rooms and 50,000 randomly selected samples for rooms with varying length and width.

## 4.2 MLP architecture

MLPs have been used for regression analysis in indoor visible light positioning systems in previous works [32, 47] where Levenberg-Marquardt (LM) is the optimizer. LM works best for shallow networks (one or two hidden layers) with a few network parameters (less than 100), it scales poorly with many neurons and needs a lot of memory. We prefer to use MLPs with low memory profile and MAE as loss function, so we chose Adam as optimizer that can work for many-layered networks with a lot of parameters.

AutoML encompasses techniques for automatically identifying the best-performing model for a specific dataset. The process involves discovering optimal model architecture and fine-tuning the hyperparameters used in training and is commonly known as neural architecture search. The open-source library AutoKeras [48] was used, that facilitates AutoML specifically for deep learning models through the TensorFlow tf.keras API, to find the MLP models that best fit the input datasets.

The input data is partitioned into three distinct subsets using a 60-20-20 ratio. Training subset is used to teach the MLP model by adjusting its weights and biases, validation subset helps assess the model’s performance during training to avoid overfitting, and testing subset shows the model's generalization capability on completely unknown data. Finally, we apply the trained model (prediction) to the dataset of values which are considered as “real values”. The input datasets for training the ML models are defined fractions of the initial datasets produced by simulation.

Starting with the single LED dataset from the mid-sized room, a search was conducted through the hyperparameter space with AutoML to determine the optimal structure for the best-performing MLP models. This search focused on identifying the ideal hidden layer (HL) count and per-layer neuron density. We also searched for the best suited optimizer and its learning rate. The input layer consists of 3 neurons (location coordinates), while hidden layers employ ReLU (rectified linear unit) activation function, and the output layer is a single neuron layer and represents the predicted RSS value.

The search ended up with two MLP models of 2 hidden layers, the first with structure 32x128 and the second with 64x256 (N1 x N2 means that first hidden layer has N1 neurons and the second N2) using the Adam Optimization algorithm with 0.001 learning rate. The MLP models are depicted in Figure 3. The input layer consists of 3 neurons for the case of a room with certain dimensions (all neurons having orange color) or 5 neurons (all neurons having blue color) for the case of rooms with length and width between 3 and 7 meters and 3m height. The input samples are normalized before being fed into the input layer. This preprocessing step ensures that the data is scaled appropriately, improving the stability and efficiency of the model during training.

## 4.3 ML models implementation

We apply these two MLP models and the DT algorithms we used in our previous work [25] to all the VLC systems. The DT algorithms are Decision Tree (DT), Extra Trees (XT), AdaBoost with Extra Trees Regressor as estimator (AdaBoost) and Stacking Regressor with four estimators which are Decision Tree Regressor, AdaBoost with Extra Trees Regressor, XGBoost Regressor and LightGBM Regressor (Stack). We conducted extensive searches of the hyperparameter space and used cross-validation to identify the optimal settings for each estimator, thereby enhancing their performance.

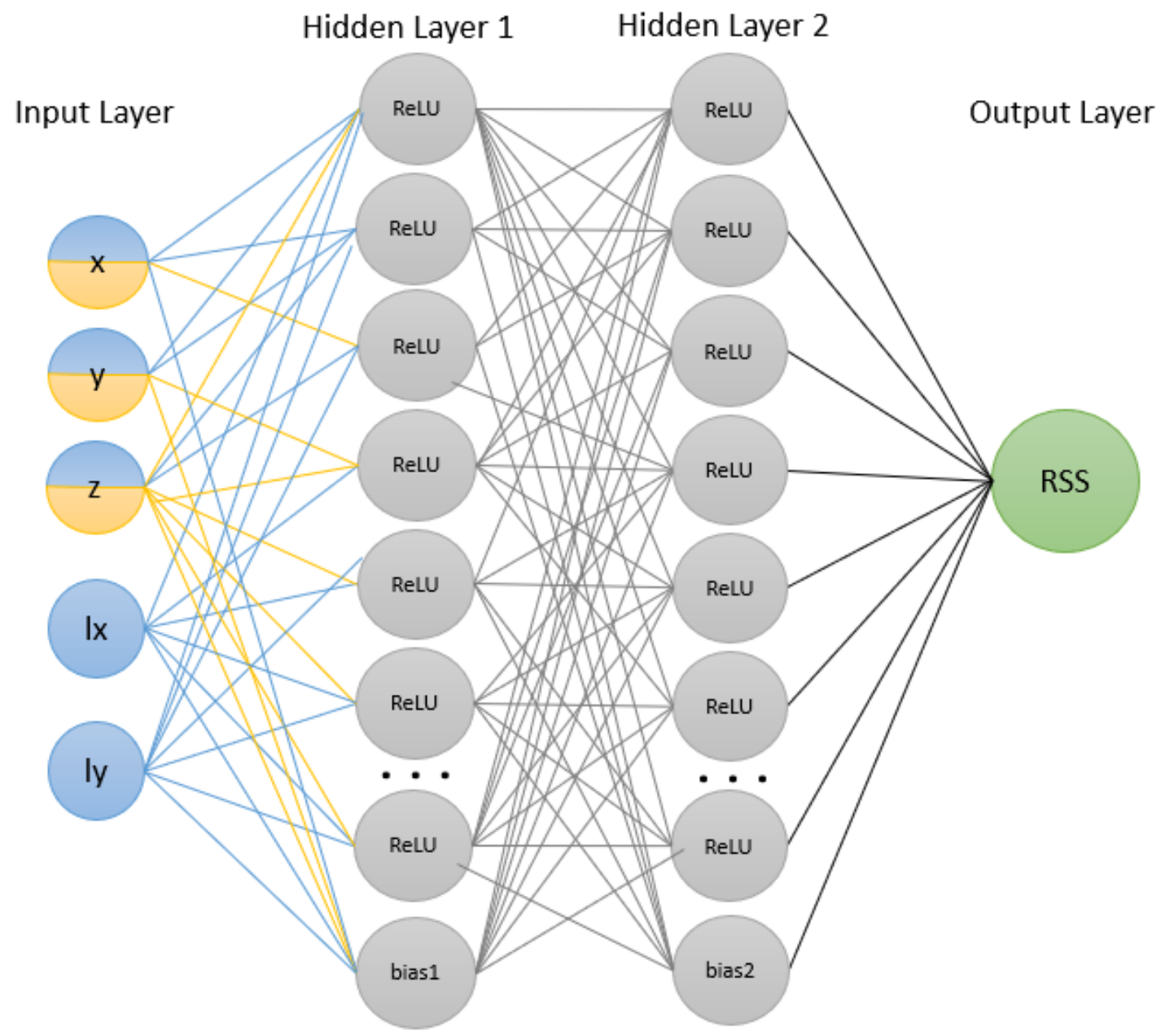

**FIGURE 3** The MLP with two hidden layers of (32×128 or 64×256) ReLU neurons for RSS prediction. Input layer for specific room (all neurons having orange color), input layer for rooms with variable length and width (all neurons having blue color).

We trained the models by performing 100 different experiments (trainings) with each model on the respective datasets to achieve statistically significant results. In each experiment, the input dataset, which is a defined fraction of the whole dataset, is randomly picked for splitting in training, validation and testing sets. The MLPs are trained over 2000 epochs with batch size of 32.

Following the approach in [25], we introduced a noise component to the RSS values in the dataset. This noise follows a Gaussian distribution $N\ (0, \sigma^2)$, where $\sigma$ is computed as the product of the standard deviation of the original RSS values and a predefined noise factor. The noise factor effectively determines the optical SNR (OSNR), which measures the ratio of the optical power reaching the receiver through an ideal (noiseless) channel to the power of the added noise. For our analysis, we used the average OSNR across all target locations in the prediction dataset, representing the room's mean OSNR level.

In Figure 4 the models' predictive accuracy for different sample size in mid-sized room with Single LED is illustrated. As expected, increasing the sample size of the training dataset leads to improved prediction accuracy as larger dataset provides the models with more information to learn from. All the VLC systems exhibit similar behavior. Indicatively, for 25k sample size MAE in dBm is depicted based on 100 experiments of the received optical power values estimated for 500 3D locations of the Rx for Four LEDs systems in small and big room (Figure 5) and for rooms with varying length and width for both LED systems (Figure 6).

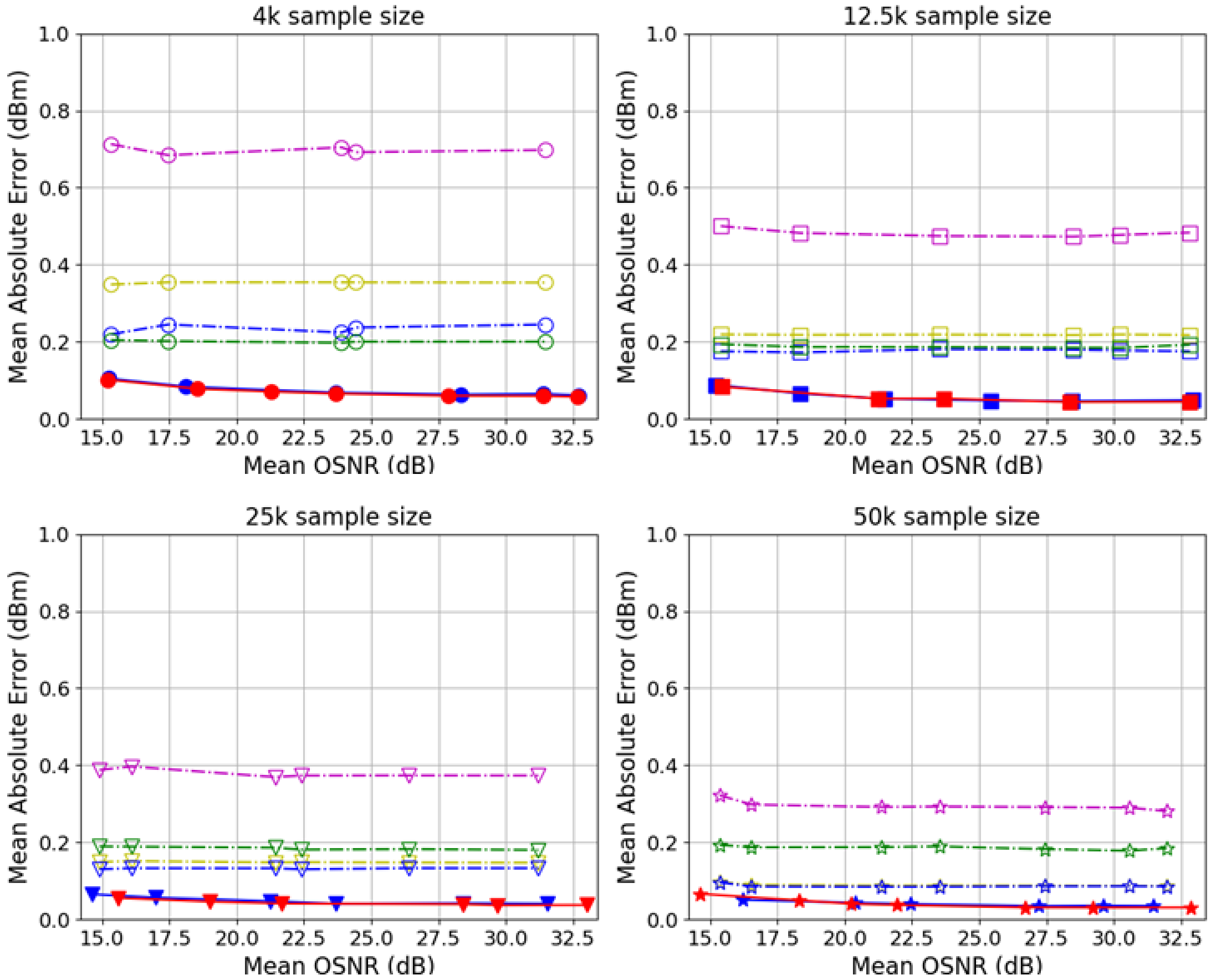

**FIGURE 4** Comparison of the ML models for different sample sizes (4k, 12.5k, 25k and 50k) randomly chosen from the 125k available in the simulation dataset in the mid-sized room with Single LED. Decision Trees (DT, XT, AdaBoost, Stack) with transparent symbols, MLPs (MLP_32×128, MLP_64×256) with filled symbols.

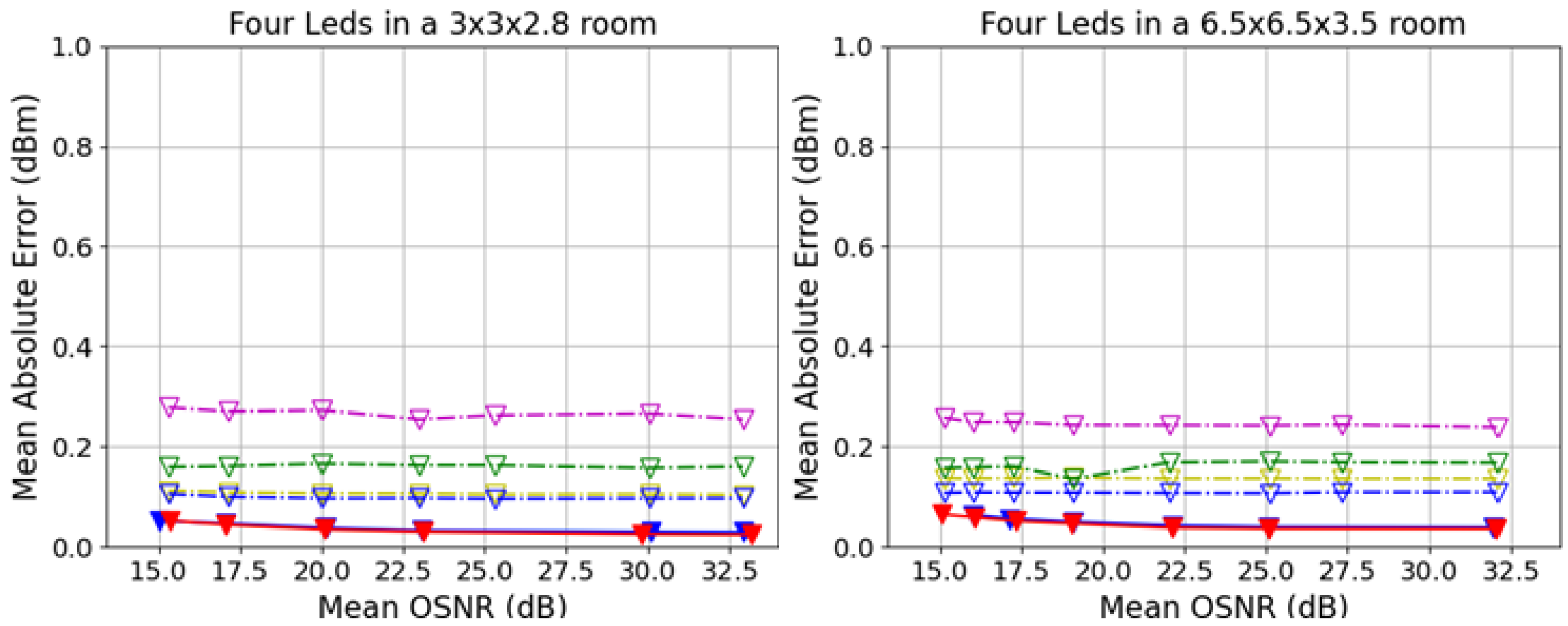

**FIGURE 5** MAE of ML models that were trained over 25k sample size dataset for the Four LEDs systems in small and big room. Decision Trees (DT, XT, AdaBoost, Stack) with transparent, MLPs (MLP_32×128, MLP_64×256) with filled symbols.

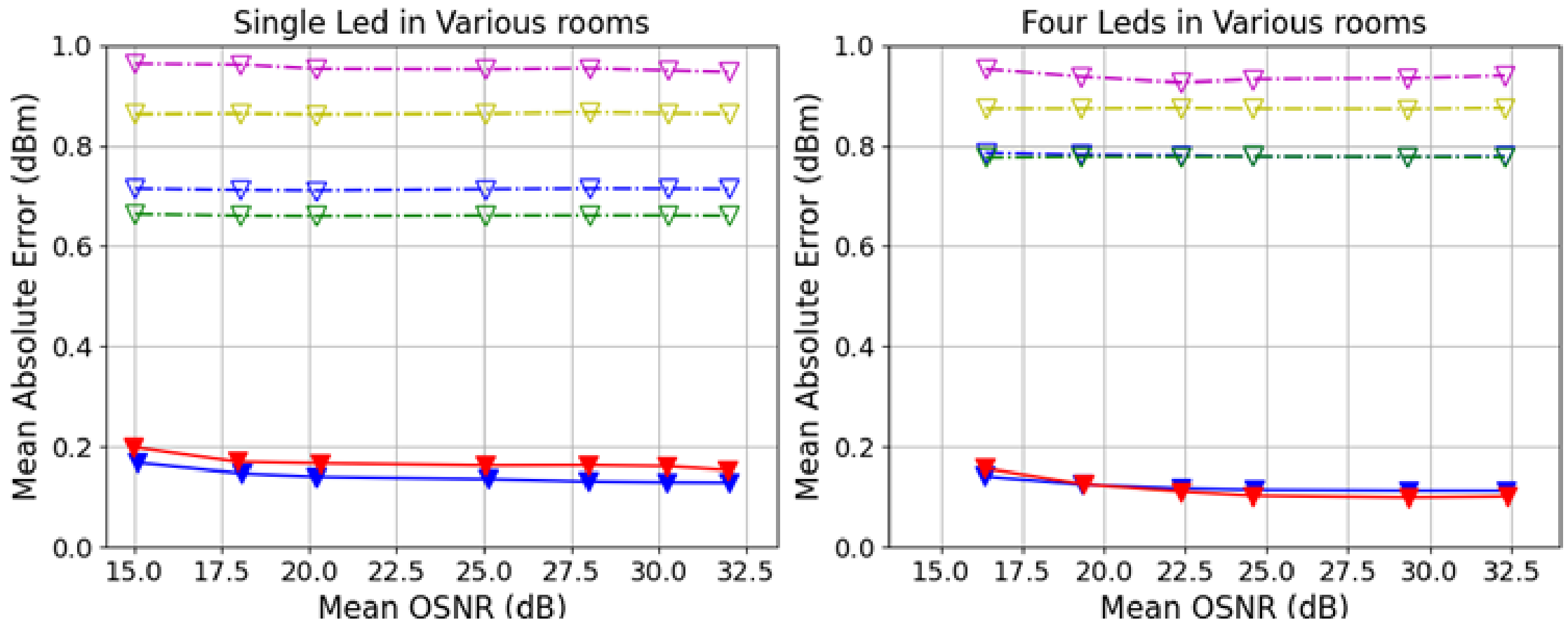

**FIGURE 6** MAE of ML models that were trained over 25k sample size dataset for both LED systems for rooms with varying length and width. Decision Trees (DT, XT, AdaBoost, Stack) with transparent, MLPs (MLP_32×128, MLP_64×256) with filled symbols.

When feature-target interactions are characterized by complex and non-linear patterns, ANNs tend to perform better as the number of features increases. In contrast, DTs often struggle to capture intricate patterns in the data with a growing number of features. This observation holds true for all VLC systems we examine. The two MLP models everywhere exhibit greater prediction accuracy than DT, and the difference is particularly pronounced in rooms with varying lengths and widths, where MLPs provide significantly more accurate predictions.

## 4.4 Balancing accuracy and efficiency of ML models

To conduct a comprehensive evaluation of ML models, we need to consider other performance metrics besides their predictive ability, such as the size of the training dataset, training time, and prediction time. This way, we will identify which models exhibit superior overall performance, balancing accuracy with computational efficiency.

### 4.4.1 Training time

First, we compare the training times of the models. We exclude the DT model due to its poor prediction accuracy, and the Stack model due to its lengthy training time. Note here that the MLP models are trained for 2000 epochs with a batch size of 32. Figure 7 illustrates MAE (averaged over 100 experiments) and the corresponding training times (averaged over 10 experiments) for XT and AdaBoost models trained on 50,000 samples, alongside MLP models trained on 4,000 and 12,500 samples, specifically for systems in the mid-sized room. XTs known for its low computational complexity, demonstrates very short training times; however, it struggles to accurately capture the intricate relationships within the VLC systems. Conversely, MLP models exhibit better or comparable accuracy despite being trained on smaller sample sizes, highlighting their efficiency in balancing accuracy with training dataset size.

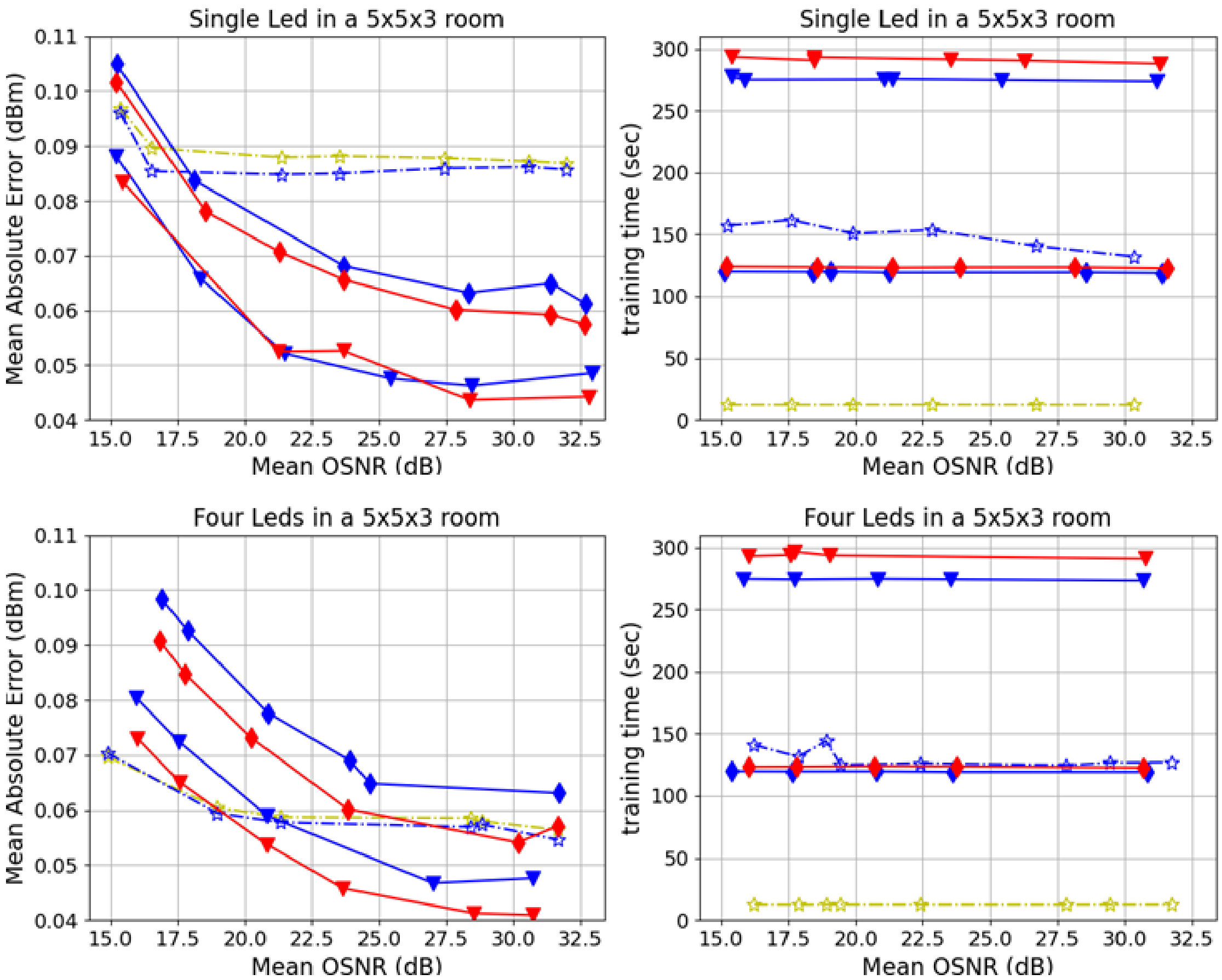

**FIGURE 7** MAE and the respective training time of XT and AdaBoost trained over 50k (✩) samples and MLPs trained over 4k (♦) and 12.5k (▼) samples for 2000 epochs with a batch size of 32. Decision Trees with transparent symbols, MLPs (MLP_32×128, MLP_64×256) with filled symbols.

So, we need to look for ways to speed up training of MLPs without significantly reducing their prediction accuracy. This can be accomplished by reducing the number of epochs. Another countermeasure is by increasing the batch size. As noted in Section 3.2.2, an epoch encompasses a single pass over the complete collection of samples, while a batch represents the predetermined number of samples processed for each parameter update iteration.

We trained our MLP models on 12.5k samples with a batch size of 32, using a reduced number of epochs—100, 250, 500, and 750—and recorded the mean training times over 100 experiments for the Single LED system in the mid-sized room. As shown in Figure 8a, it logically follows that longer training results in better prediction accuracy. However, training times comparable to those of the XT algorithm trained for 50k samples, were observed when the models were trained for 250 epochs or fewer. Next, we trained our MLP models for 250 epochs with varying batch sizes—32, 64, 128, and 192—and recorded the mean training times over 100 experiments. As shown in Figure 8b, using a batch size of 128 allows the MLPs to achieve low training times while still maintaining superior prediction accuracy compared to the XT model.

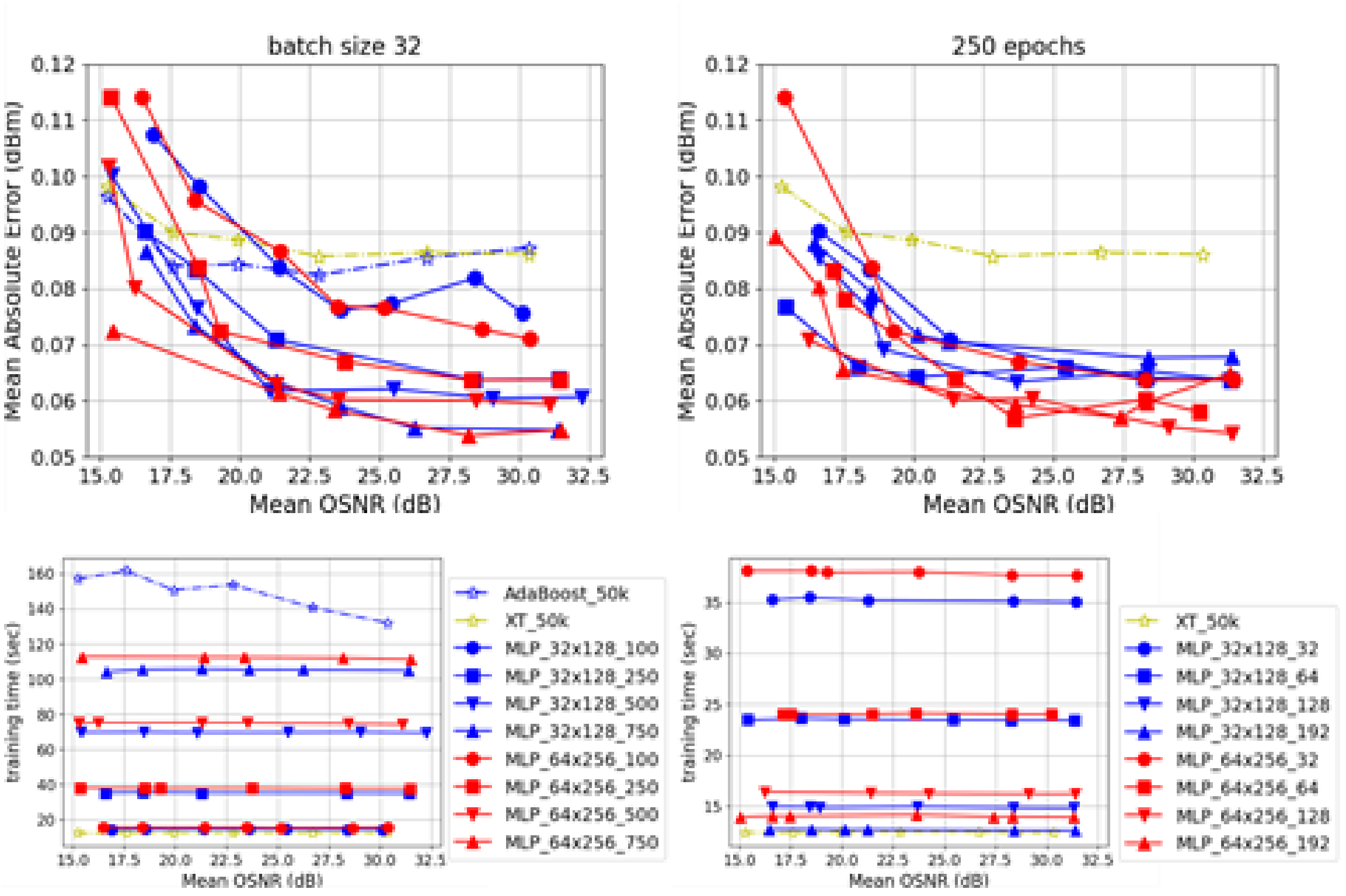

a) b)

**FIGURE 8** MAE and the respective training time of a) XT and AdaBoost trained over 50k samples and MLPs trained over 12.5k samples for batch size 32 and different number of epochs b) XT and MLPs trained over 12.5k samples for 250 epochs and different batch sizes.

Focusing on models with low prediction times, Figure 9 illustrates MAE and training times, both averaged over 10 experiments, exclusively for the XT and MLP models trained for 250 epochs with a batch size of 128. All models were trained on sample sizes of 4k, 12.5k, and 25k. To highlight the low predictive capability of the XT model, we also include its MAE when trained on the entire dataset of 400k samples. The results clearly show that MLPs, even when trained on a smaller sample size of 4k, can achieve training times comparable to those of the XT model while delivering significantly better predictions.

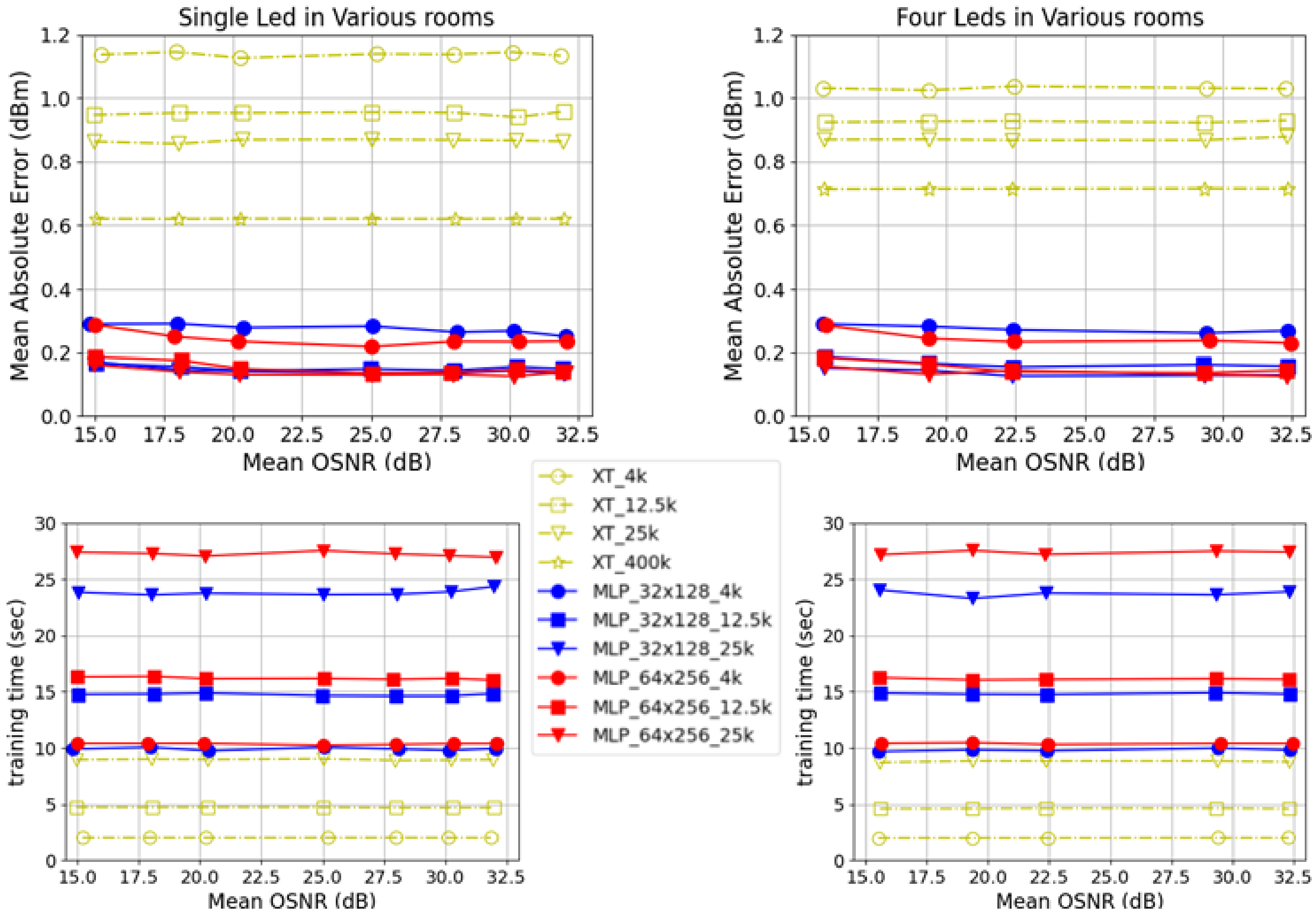

**FIGURE 9** MAE and training time for XT and MLPs trained for 250 epochs with a batch size of 128, for different sample sizes.

In fact, parameters of MLP models, such as sample size, the number of epochs and batch size, can be adjusted to balance the desired level of inference accuracy with training time. This flexibility is particularly valuable for ML models that require frequent retraining or are integrated into real-time systems like Digital Twins, where minimizing training time is critical. By fine-tuning these parameters, one can optimize the model's performance to meet specific operational needs.

The reason why the two MLPs with different hidden layer sizes take the same time to train for the same sample size could be due to how efficiently the CPU and TensorFlow handle the computations. Modern CPUs with many cores and GPUs are built to perform large parallel tasks effectively. If the batch size is large, both networks can fully use the processor's power, making their training times similar. The time per epoch may be more about managing data batches than the computations within the layers. TensorFlow also optimizes the computational graph, potentially leading to similar training times if it optimizes both networks equally and executes operations in parallel.

Figure 10 shows the distribution of MAE over 100 experiments for 50k reference values for the one and four LED system in varying size rooms, for the two MLP models trained for 250 epochs with 128 batch size using randomly picked samples of 12.5k out of 400k in each experiment. In a violin plot, wider sections indicate higher probability density. Within the plot, white dot marks the median value, and the central black bar shows the dispersion from the first to the third quartile. As SNR increases the outliers are less, the median decreases and its probability increases. The standard error of the mean is 0.002 dB for mean OSNR values greater than 20 dB.

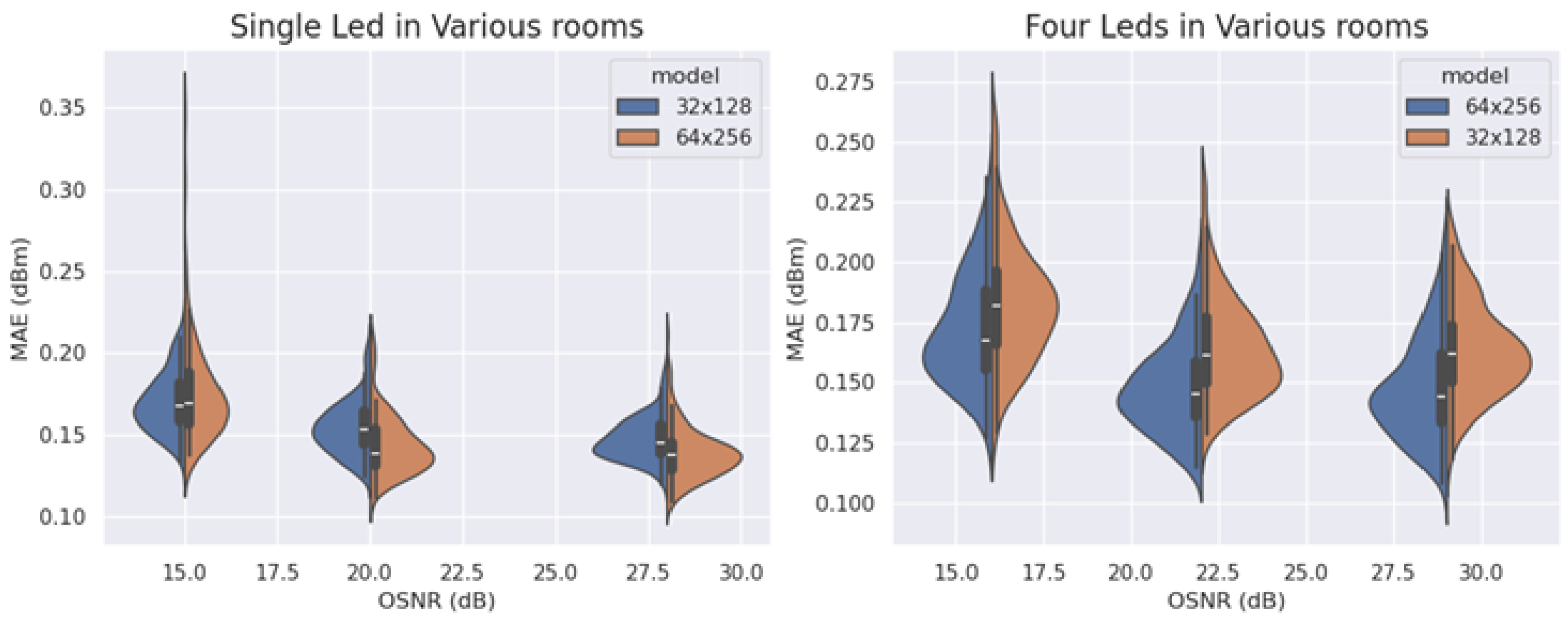

**FIGURE 10** Distribution of MAE over 100 experiments for 50k reference values for the two MLP models trained for 250 epochs with 128 batch size using randomly picked samples of 12.5k out of 400k in each experiment.

The highest light intensity is observed directly beneath the LED source(s), gradually decreasing as one move toward the walls and corners of the room. This is clear in the graphs of Figure 11, where the OSNR on the 50 locations for prediction along the half-diagonal on a plane 1m above the floor in mid-sized room are shown. The reliability and predictive effectiveness of the MLP algorithms, 32x128 and 64x256 trained for 250 epochs with 128 batch size, are clearly demonstrated in the Figure, showcasing their superior performance in predicting outcomes compared to DT models.

#### 4.4.2 Prediction time

Except training time, prediction (inference) time is equally important, especially in real-time applications where quick predictions are required. Table 2 presents prediction time for XT models and MLPs for a single location. AdaBoost is not referenced as it needs over 700 μsec to predict a single value.

**TABLE 2** Prediction Time of ML Models Trained Over 25k Sample Size.

| **Room** | **Single LED** | | **Four LEDs** | |
|---|---|---|---|---|
| | **XT (μsec)** | **MLPs (μsec)** | **XT (μsec)** | **MLPs (μsec)** |
| Small | 90 | 180 | 90 | 180 |
| Mid-sized | 90 | 180 | 90 | 180 |
| Bigger | 90 | 180 | 90 | 180 |
| Varying size | 26 | 21 | 21 | 21 |

All calculations for timing recordings were executed on a workstation configured with a 4-core Intel(R) Xeon(R) E-2224 CPU @ 3.40GHz and 80GB of DDR4 RAM operating at 2666 MT/s.

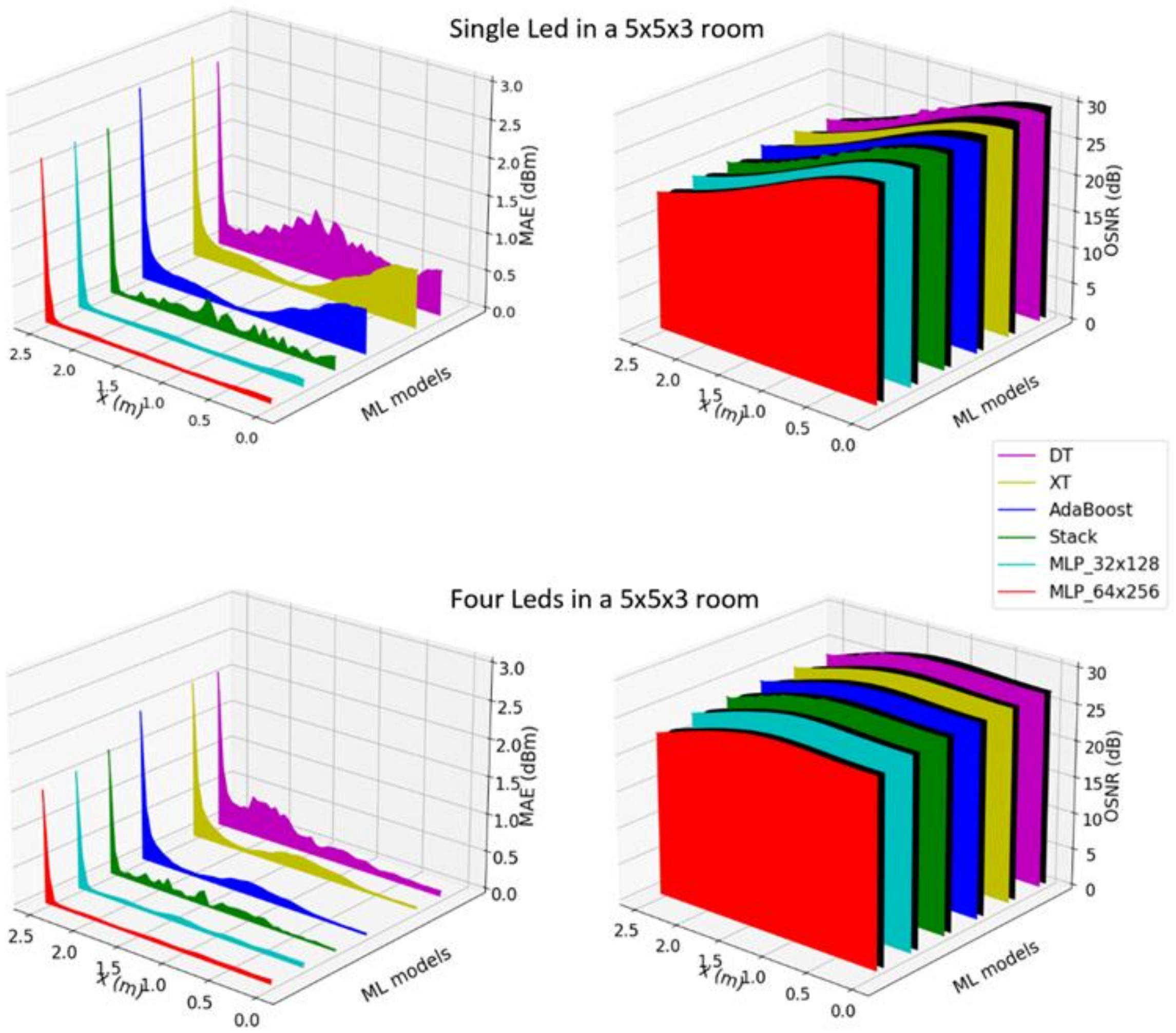

**FIGURE 11** MAE and OSNR averaged over 100 experiments for Single and Four LEDs system in mid-sized room along the half-diagonal (x is the abscissa of the points on the diagonal) on a plane 1m above the floor. The sample size is 12.5k out of 125k for all models. Black graphs are the reference values for OSNR.

## 5 MLP Evaluation on Experimental Data

We perform a comprehensive comparison on experimental data of the proposed MLP model through benchmarking, spatial validation, and robustness analysis to evaluate its predicting performance and generalization ability across the entire spatial domain. Specifically, we compare the MLP (64×256) with a set of representative approximation methods in RSS spatial modeling. This set includes Inverse Distance Weighting (IDW) and Thin Plate Splines (TPS) as basic deterministic interpolation methods, and Gaussian Process Regression (GPR) and Support Vector Machines (SVM) with RBF kernels as basic kernel-based regression methods. We also include representatives from other neural network architectures such as Extreme Learning Machines (ELM), Convolutional Neural Networks (CNN) and U-Net architectures in the set of methods. Finally, we add a hybrid ensemble that combines from the above neural models, MLP, CNN and U-Net, to evaluate whether the increased complexity of the model yields measurable performance gains. The evaluation protocols and results are described in detail in Sections 5.2–5.7.

## 5.1 Experimental VLC dataset and sampling

The data we use to evaluate the MLP model originates from the Public VLP Dataset repository [49], corresponding to the work of Glass et. Al [50] The dataset covers 7344 RSS measurements with the photodetector on the same plane (a few cm above the floor) over a 6.3 × 6.9 m test area illuminated by 11 VLC luminaires.

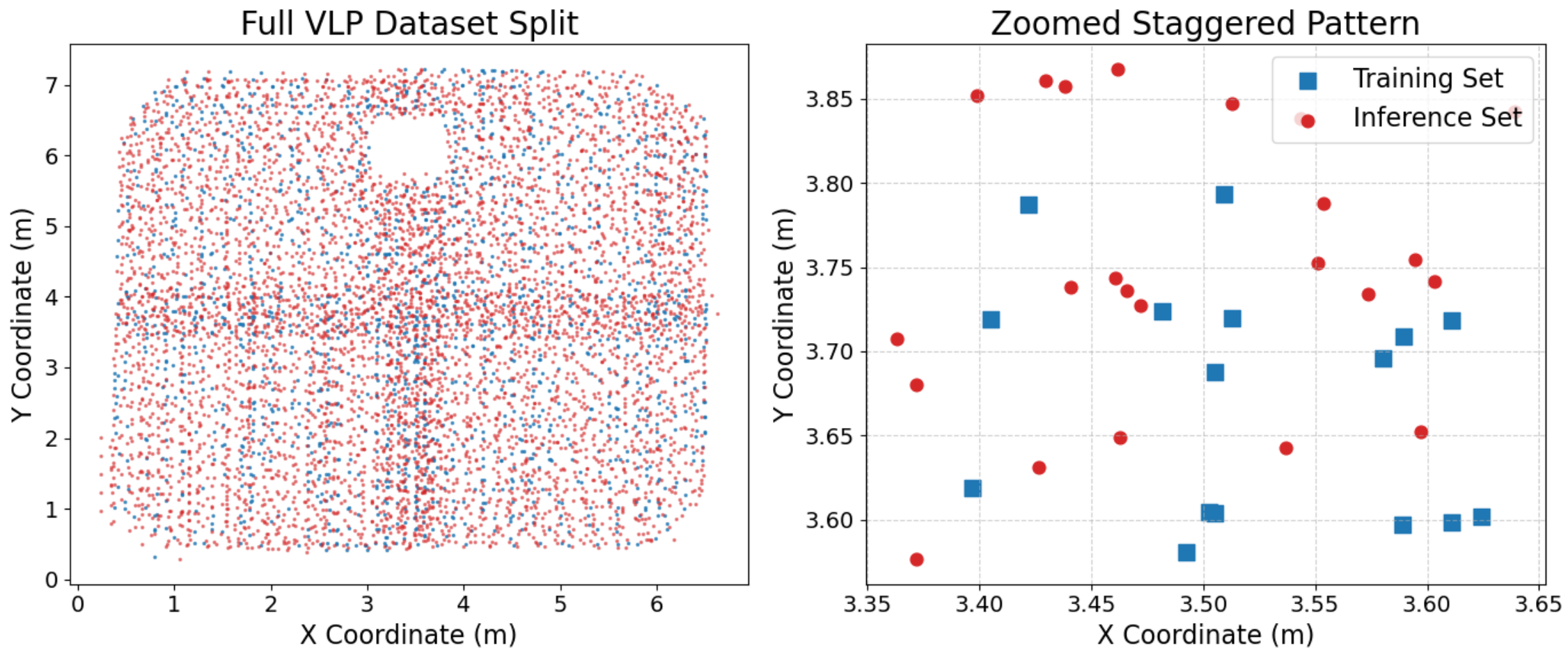

**FIGURE 12** Spatial layout of the staggered grids partitioning comprising training (blue) and inference (red) samples in the Public VLP experimental environment.

We selected the deterministic sampling strategy of staggered grids to prevent optimistic bias often caused by spatial correlation. In the fields of geostatistics and signal processing, staggered (or hexagonal) grids are widely recognized as the optimal geometry for two-dimensional sampling. Olea showed [51] that such an arrangement minimizes the maximum distance from any unsampled coordinate, and thus inherently reduces spatial prediction variance. We applied this sampling scheme to the 7344 experimental measurements and created two spatially interleaved and completely non-overlapping grids, to ensure that no inference location ever coincides with a training point. The first grid consists of 1818 samples for training and the second with the remaining 5526 strictly for inference (see Figure 12 with blue and red color). All evaluations described in Section 5 were performed using this fixed partitioning, with the predictive performance being quantified mainly through MAE and averaged over one hundred (100) independent runs.

## 5.2 Models' configuration and hyperparameters

To ensure a fair comparison of CNN and U-Net networks with MLP in terms of model capacity, the CNN and U-Net architectures were configured to match the hidden layer widths of MLP (64 and 256). They were also adjusted to the tabular representation of the samples. The MLP processes the input feature vector using fully connected layers with global receptive fields. Although distance features encode geometric information, the input remains tabular and doesn't form a structured spatial grid, so it is evaluated point by point under the staggered grid protocol. Consequently, two-dimensional convolutions and spatial pooling are not applicable. Instead, one-dimensional convolutions are employed, where the CNN uses 64 and 256 convolutional filters to operate along the feature dimension. A conventional U-Net with downsampling cannot be directly applied in this setting; therefore, a skip connection convolutional architecture is implemented, that features a 64→256 filter bottleneck and direct concatenation of the input features with deeper representations prior to the final prediction. This design preserves the spirit of U-Net style feature reuse and remains compatible with point wise regression. All neural networks were trained for 250 epochs with a batch size of 128, values of the specific training parameters that, as we showed in section 4.4, give a balanced performance between prediction accuracy and training time in our MLP model.

GPR is an optimal spatial interpolator when trained locally on dense data (as utilized for synthetic data generation), its $O(N^3)$ computational complexity renders it fundamentally impractical for global modeling of large-scale multi-transmitter environments. To enable GPR effectively scale and use 100% of the samples information without the $O(N^3)$ memory crash we use K-Means clustering to generate inducing points that are representative across the entire room.

Input feature vector comprising receiver coordinates and distances to LED transmitters while targets of models are power received from each LED individually. The resulting model configurations and key hyperparameters for all methods are summarized in Table 3.

**TABLE 3** Models' Configuration Summary.

| Method | Structural Characteristics | Key Parameters & Hyperparameters |
|---|---|---|
| MLP | Processes tabular point-wise features using global receptive fields | 2 hidden layers (64 × 256 nodes), ReLU, Adam, LR = 0.001 |
| CNN | No 2D spatial pooling; uses 1D convolutions operating strictly along the feature dimension | 1D conv layers (64 & 256 filters), kernel size = 2, ReLU, Adam, LR = 0.001 |
| UNet | Adapts U-Net style feature reuse to tabular data by concatenating input features with deep representations | 1D Encoder–decoder, skip connections, 64 → 256 filter bottleneck, kernel size = 2, ReLU |
| SVM | Standard Support Vector Regression mapped to a higher-dimensional space | RBF kernel, C = 100, γ = 'scale' |
| GPR | Uses K-Means clustering to generate 'inducing points' to scale the GPR memory efficiently | RBF kernel, length scale = 1.0, 400 inducing points |
| TPS | Radial Basis Function (RBF) interpolator for multidimensional point-wise data | Thin-plate spline kernel, Smoothing parameter λ = 0.01 |
| IDW | Traditional spatial interpolation based on exact physical proximity | Power parameter p = 2 |
| ELM | Single-hidden-layer feedforward network with fixed random hidden nodes; only output weights are learned | 512 hidden nodes, random input weights, ReLU activation |
| Hybrid | Ensemble of the deep learning architectures MLP, CNN 1D, and UNet 1D | weights optimized via Sequential Least SQuares Programming (SLSQP) |

## 5.3 Benchmarking under staggered-grid sampling

In Table 4 we present the prediction accuracy of all selected methods, under staggered grid sampling, where the training and inference points are spatially interleaved but non-overlapping. We observe that the MLP, CNN-1D, U-Net-1D and TPS methods achieve almost identical MAE values between 0.24-0.25. This fact shows that the underlying RSS field can be accurately approximated by a variety of regression approaches, either neural models or classical interpolation methods, when the inference locations are close to the observed training points.

In contrast, methods that rely on stronger global assumptions or have limited spatial adaptability, such as GPR and IDW, perform significantly worse. In particular, the lower performance of GPR suggests that without extensive tuning, kernel-based regression is not well suited to the scale or noise characteristics of the real data set we use.

**TABLE 4** Prediction accuracy under staggered-grid sampling.

| Model | MAE (dBm) |
|---|---|
| MLP (64×256) | 0.25 |
| CNN-1D (64×256) | 0.24 |
| UNet-1D (64×256) | 0.24 |
| SVM (RBF) | 0.30 |
| ELM (512 nodes) | 0.30 |
| GPR | 3.58 |
| IDW (p = 2) | 0.83 |
| TPS | 0.24 |
| Hybrid (MLP+CNN+UNet) | 0.24 |

## 5.4 Spatially Blocked Cross-Validation

With spatially blocked cross-validation we evaluate the spatial generalization ability of the models by requiring them to extrapolate to entire spatial regions that were not included during their training. To achieve this, we divide the room into 4 quadrants, train models on 3, and test on the 1 unseen quadrant cyclically.

As shown in Table 5, convolutional architectures exhibit the best robustness of all models. This suggests that convolutional inductive bias provides some advantage in predicting RSS in unseen regions, perhaps by exploiting local spatial regularities that are transferable across the room. The hybrid ensemble achieves performance comparable to convolutional models, but without a clear advantage. Classical interpolation methods, as well as kernel-based ones, degrade sharply, with GPR and IDW exhibiting MAE values exceeding 3dBm. Overall, spatial generalization remains poor and spatial extrapolation remains difficult regardless of the model architecture.

**TABLE 5** Spatial generalization performance under blocked cross-validation (unseen regions).

| Model | Mean MAE ± Std (dBm) |
|---|---|
| MLP (64×256) | 0.80 ± 0.04 |
| CNN-1D (64×256) | 0.62 ± 0.02 |
| UNet-1D (64×256) | 0.64 ± 0.02 |
| SVM (RBF) | 1.09 ± 0.00 |
| ELM (512 nodes) | 2.61 ± 0.16 |
| GPR | 4.65 ± 0.09 |
| IDW (p = 2) | 3.26 ± 0.00 |
| TPS | 1.00 ± 0.00 |
| Hybrid (MLP+CNN+UNet) | 0.65 ± 0.02 |

## 5.5 Model Capacity Ablation

We performed model capacity ablation exclusively for the MLP, which is the main model under investigation. Table 6 lists the impact of the MLP capacity on prediction accuracy and computational cost. We tested two-hidden layer configurations of shallow, medium, and deep architectures, as well as a three-hidden layer configuration. The MAE remains constant at approximately 0.25dBm, with overlapping 95% confidence intervals. For all configurations, the training time increases moderately with network size, while the inference time remains essentially constant. These findings support the use of the medium-sized selected MLP since increasing capacity does not provide any tangible benefit.

**TABLE 6** Effect of MLP capacity on performance and runtime.

| MLP Configuration | MAE (95% CI) | Training Time (s) | Inference Time (s) |
|---|---|---|---|
| Small (32×128) | 0.25 [0.24, 0.25] | 5.60 ± 0.11 | 0.20 ± 0.00 |
| Medium (64×256) | 0.25 [0.24, 0.26] | 6.48 ± 0.34 | 0.20 ± 0.00 |
| Large (128×512) | 0.25 [0.25, 0.27] | 7.32 ± 0.29 | 0.20 ± 0.00 |
| 3 HLs (32×64×128) | 0.25 [0.25, 0.26] | 5.77 ± 0.12 | 0.20 ± 0.00 |

## 5.6 Robustness to Input and Label Noise

Robustness results under input perturbations and label noise are summarized in Table 7. When noise is introduced at the receiver location, the ML models show gradual degradation, with MAE increasing as the magnitude of the perturbation increases. MLP, CNN-1D and U-Net-1D neural networks show similar behavior, except for the ELM which fails catastrophically even for small perturbations. SVM also exhibits similar behavior to the neural networks, TPS appears insensitive to jitter, while the classical interpolation methods show limited tolerance. Finally, GPR performs poorly at all noise levels.

In label-noise tests, RSS labels are corrupted by additive noise, simulating measurement errors and reduction in signal-to-noise ratio (SNR). As the SNR decreases, all models show performance degradation. MLP degrades more slowly than convolutional architectures, SVM remains competitive, while TPS and ELM show rapid error growth. These results indicate that increased model complexity enhances sensitivity to label noise, while simpler architectures show greater robustness.

**TABLE 7** Robustness of selected models to sensor jitter and label noise.

| **Model** | **Sensor Jitter (Distance Noise)** | | | **Injected Label Noise Standard Deviation $\sigma$ (Low-SNR Training Data)** | | |
|---|---|---|---|---|---|---|
| | **0.2 m** | **0.5 m** | **1.0 m** | **2 dBm** | **5 dBm** | **10 dBm** |
| MLP (64×256) | 0.49 | 1.03 | 1.89 | 0.47 | 0.88 | 1.77 |
| CNN-1D (64×256) | 0.51 | 1.11 | 2.09 | 0.55 | 1.28 | 2.73 |
| UNet-1D (64×256) | 0.51 | 1.10 | 2.05 | 0.58 | 1.38 | 3.08 |
| SVM (RBF) | 0.42 | 0.88 | 1.98 | 0.44 | 0.87 | 1.67 |
| ELM (512 nodes) | 7.88 | 19.76 | 39.41 | 1.20 | 2.87 | 5.71 |
| GPR | 3.69 | 3.89 | 3.93 | 3.29 | 3.72 | 3.87 |
| IDW (p = 2) | 1.49 | 2.13 | 2.76 | 0.97 | 1.42 | 2.32 |
| TPS | 0.24 | 0.24 | 0.24 | 0.89 | 2.11 | 4.19 |
| Hybrid (MLP+CNN+UNet) | 0.64 | 1.41 | 2.63 | 0.46 | 0.89 | 1.75 |

## 5.7 Runtime Analysis

In this section the computational costs of all evaluated methods are reported and compared (Table 8), including the physics-optimized VLC Room Simulator described in Section 2.2. Training time for the Simulator refers to a pre-compute calculation of room environment where the Array-based Radioactivity Method is applied and the wall-to-wall reflection matrix multiplications are timed. Also, the single-point and full-grid times only measure the time to compute the LOS and NLOS paths for all reflections with a maximum of 10 bounces on walls, floor, and ceiling.

Among ML models, MLP offers a favorable balance between training time and inference speed, while convolutional architectures have significantly higher training costs without faster inference. The hybrid ensemble, despite its competitive accuracy, has a long training time, since each of the constituent models needs to be trained separately.

**TABLE 8** Training and inference runtime comparison of REM estimation methods.

| Model | Training Time (s) | Grid-Level Inference (s) | Single-Point Inference (ms) |
|---|---|---|---|
| MLP (64×256) | 6.36 | 0.19 | 30.9 |
| CNN-1D (64×256) | 21.78 | 0.27 | 30.9 |
| UNet-1D (64×256) | 34.10 | 0.33 | 30.7 |
| SVM (RBF) | 2.12 | 0.56 | 1.9 |
| ELM (512 nodes) | 1.02 | 0.15 | 1.0 |
| GPR | 6.19 | 0.53 | 2.1 |
| IDW (p = 2) | 0.00 | 1.31 | 1.2 |
| TPS | 0.55 | 1.34 | 1.5 |
| Hybrid (MLP+CNN+UNet) | ~ 63 | 0.56 | 91.6 |
| Optimized VLC Room Simulator | 10.86 | 3.60 | 94.8 |

Calculations for timing recordings were executed on the same workstation referenced in section 4.4.2.

All the above results of benchmarking, spatial validation and robustness analysis of the selected models highlight the practical advantage of data-driven REM approaches, which can offer comparable or superior performance at a fraction of the computational cost required by a physics-based simulator. Overall, MLP provides the most favorable balance between accuracy, spatial generalization, robustness and computational efficiency, supporting its suitability as a practical basis for VLC REM estimation.

## 6 Synthetic Data Generation and Evaluation

To collect dense measurement data is often expensive, time-consuming and impractical. The use of synthetic data to augment a limited number of real measurements can compensate for this fact. In this section we explore the generation and use of synthetic RSS data to improve model training under reduced measurement availability. We leverage the Public VLP experimental dataset as a reference and examine oversampling-based statistics and synthesis methods to assess their effectiveness when combined with progressively smaller subsets of real training data. We evaluate the performance of the methods exclusively on real measurements according to the subsampling method described in 6.1, thus performing a rigorous evaluation of synthetic to real generalizability.

## 6.1 Subsampling Strategy from the Experimental Dataset

To simulate scenarios where only sparse training measurements are available and systematically evaluate how synthetic data generation can compensate for limited real-world observations., we extract progressively smaller subsets from the first staggered grid comprising 1818 experimental training measurements. On each subset we apply the selected synthesis methods to generate synthetic RSS samples. Then, the models are trained on the set comprising of the subset's real samples and the corresponding synthetic data derived from them. We perform the final evaluation on the second staggered grid of 5526 real measurements, which remains completely invisible during both real data selection and synthetic generation. We follow this procedure to evaluate the impact of synthetic augmentation under strict spatial separation and without information leakage.

Data in the staggered grid are sorted by Tx's coordinates, so we achieve uniform downsampling by systematically keeping every N-th sample in the sorting order. Specifically, to obtain one third of the training data (N=3, 606 samples) every third grid point is selected, to obtain one sixth (N=6, 303 samples) every sixth grid point is selected and so on. In this way, we reduce the number of available training samples while maintaining the sampling strategy of staggered grids, as described in section 5.1.

We filter the 5526 samples of the inference set for points on or near (with tolerance of 10cm) the diagonal of the 6.3×6.9m test area. Then we plot the actual and predicted total power (dBm) received at each location along the length of diagonal. In Figure 13a we see how well the MLP 64x256 model trained on the 1818 grid samples predicts the (blue) ground truth values of the 5526 grid inference set. In b, c and d of Figure 13 we see predictions of the model trained on grids comprised 101, 303 and 606 samples, respectively (as described above) on the same ground truth values. It is clear, that increasing the number of samples of the training set causes a gradual decrease in the prediction error, which is more pronounced in the corners of the room.

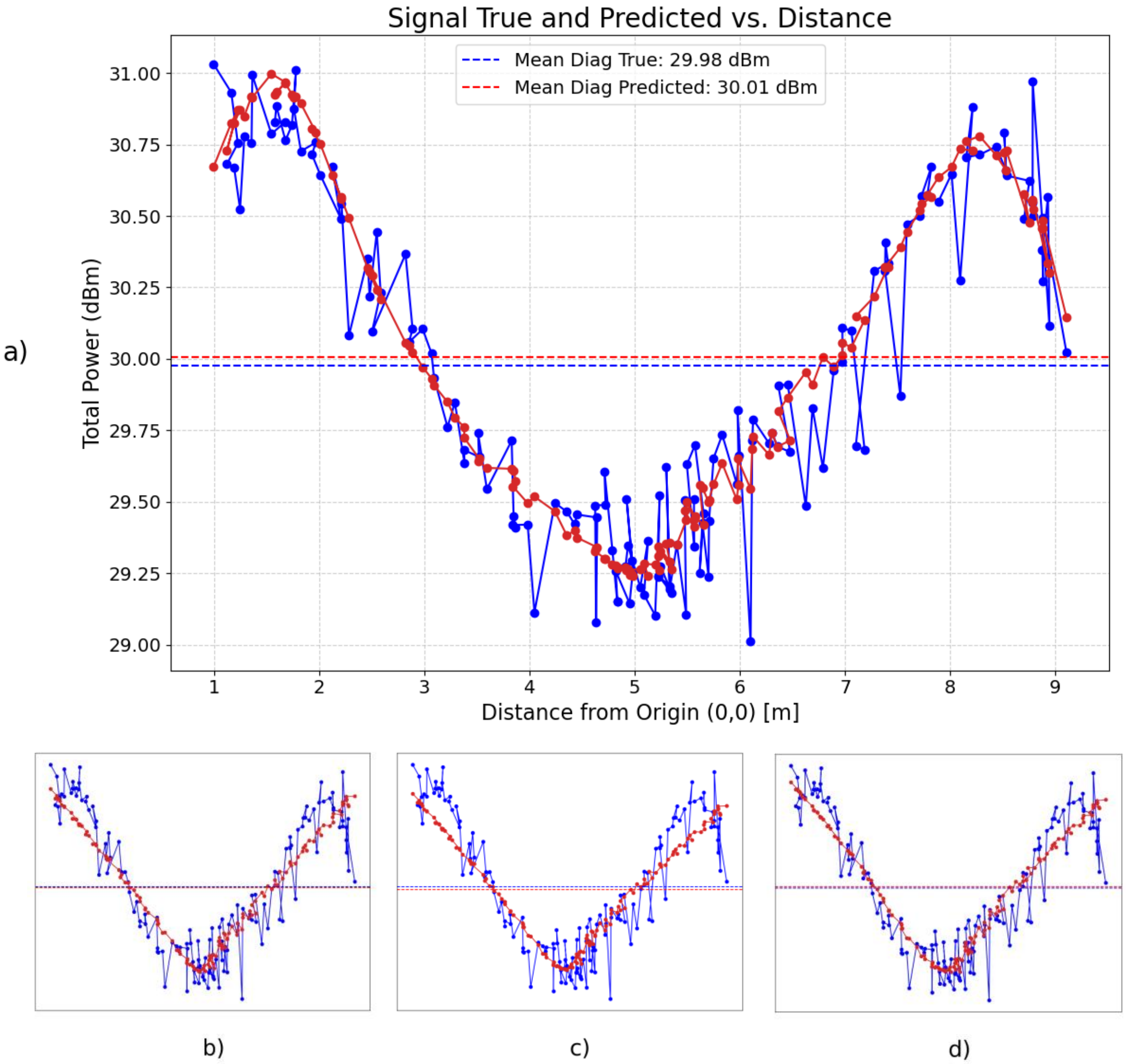

**FIGURE 13** Ground truth (blue) of 5526 experimental values of second staggered grid and predicted (red) total power (dBm) received at each location along the length of diagonal for MLP trained on a) 1818 first staggered grid samples b) subset of 101 samples c) subset of 303 samples d) subset of 606 samples

## 6.2 Synthetic RSS Data Generation Methods

The adoption of synthetic data generation has proven essential in overcoming data sparsity challenges within the field of wireless positioning, where collecting extensive real-world measurements is often labor-intensive, time-consuming, and costly. Methods for generation of synthetic data create entirely new data by learning the underlying distribution. In the context of RSS-based indoor positioning, several well-established approaches have been proposed in the literature. Physics-based methods leverage propagation models such as the log-distance path loss model combined with log-normal shadowing to generate realistic RSS values that conform to electromagnetic wave propagation principles [52, 53]. Statistical interpolation techniques, including Gaussian Process Regression (GPR) and Kriging, model the spatial correlation structure of RSS measurements and use the model's posterior to generate representative new samples [54, 55]. Deep generative models, such as Variational Autoencoders (VAEs) and Generative Adversarial Networks (GANs), have been applied to learn complex RSS distributions directly from data, though they typically require larger training datasets to avoid overfitting [56, 57]. More recently, over-sampling techniques originally developed for imbalanced classification problems have been adapted for regression tasks in wireless localization. The Synthetic Minority Over-sampling Technique (SMOTE) [58] generates synthetic samples by interpolating between existing data points within the feature space. Specifically, it populates the minority class by generating novel instances along the vector paths that link a given data point to its k-nearest neighbors. SMOTE has been extended to regression problems through (SMOTER) [59] and has been refined as SMOGN [60], that combines SMOTE-based interpolation with Gaussian noise injection to handle the unique challenges of continuous target variables while preserving local data distribution characteristics.

### 6.3 Analysis of Synthetic–to–Real Distribution Matching

In this study we choose SMOGN and GPR as synthetic RSS data generating methods. We compare synthetic datasets created by the above two generators against real RSS distribution using first order statistics and distribution level tests. Table 9 presents the statistical comparison between the real RSS measurements on the inference staggered grid (5526 points) and synthetic training datasets generated from the subsampled training set comprising 606 samples, which represents a moderate level of measurement sparsity and allows clearer inspection of the behavior of the synthetic data generators.

In indoor VLC systems, like the one under study, we know that weak RSS values correspond to receiver locations far from the LED emitters, usually near the edges and corners of the room. We encode this knowledge by assigning a high relevance score to low RSS values, moderate relevance to mid-range RSS values, and zero relevance to the strongest signals. The distance-based SMOGN method we use in this study relies on the above affinity mapping to explicitly define which samples are considered important for generating synthetic data. We apply a relevance threshold and restrict the generation of synthetic samples exclusively within the lowest quartile of the RSS distribution, i.e. for weak RSS signals far from the LEDs. In this way, we guide the SMOGN method to perform targeted oversampling only in minority spatial regions, without increasing the remaining samples.

For the case of 606 samples, SMOGN with a relevance threshold of 0.6 produces 172 additional synthetic samples. To make a fair comparison between the two generators, we chose to generate a 13×13 grid of synthetic samples from GPR so that the synthetic datasets produced by the two methods would be of similar volume.

Results reported in Table 9 and depicted in Figure 14 highlight the fundamentally different goals of the two synthesis approaches. GPR emphasizes global statistical fidelity, generating synthetic data that closely matches the overall RSS distribution. SMOGN, in contrast, emphasizes minority space enrichment and deliberately reshapes the training distribution to improve the representation of weak signal regions.

**TABLE 9** Statistical comparison and distribution tests of synthetic vs experimental data.

| Metric | Real | SMOGN Synthetic | GPR Synthetic | SMOGN Error (%) | GPR Error (%) |
|---|---|---|---|---|---|
| Mean | 29.37 | 27.85 | 29.36 | 5.17 | 0.06 |
| Std Dev | 1.44 | 1.30 | 1.54 | 9.61 | 6.86 |
| Min | 24.18 | 24.21 | 24.61 | | |
| Max | 31.80 | 29.23 | 31.51 | | |
| Median | 29.65 | 28.25 | 29.73 | | |
| **Distribution Tests** | | | | | |
| KS Stat | | 0.66 | 0.05 | | |
| KS p-val | | 0.00 | 0.88 | | |
| Wasserstein Distance | | 1.52 | 0.10 | | |

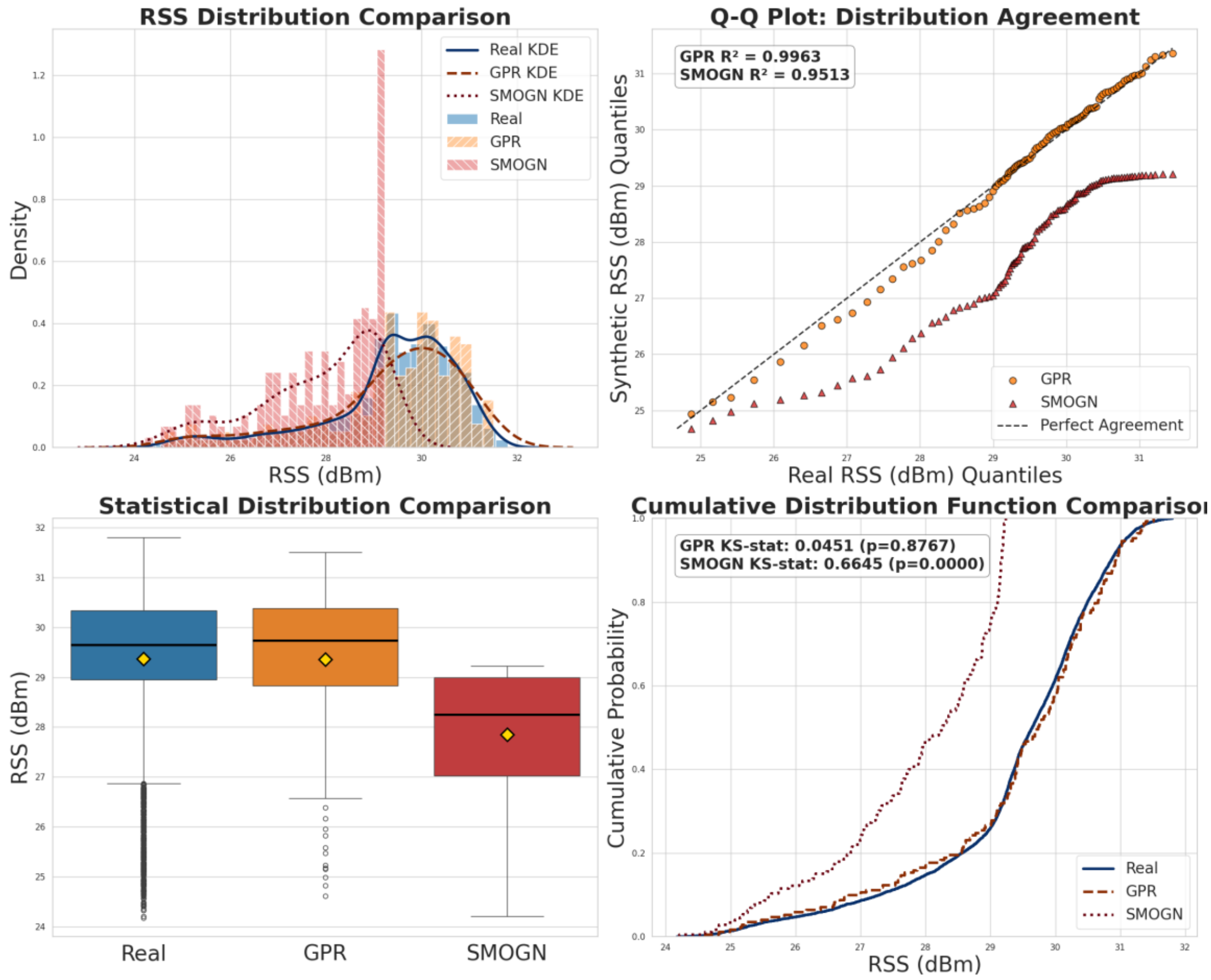

**FIGURE 14** SMOGN and GPR synthetic training data generated from 600 real values subsampled from the 1818 staggered grid compared with real 5526 staggered grid VLC RSS measurements. (a) RSS distribution with kernel density estimation, (b) Q-Q plot for quantile agreement (c) box plot comparison, (d) CDFs.

## 6.4 Training on Synthetic Data and Evaluation on Real Measurements

Using the subsampled training sets defined in Section 6.1, training is performed using combinations of real and synthetic RSS samples derived from the first staggered-grid subset. These subsets represent different levels of measurement sparsity and serve as the sole source for synthetic data generation.

For each selected subset, synthetic RSS samples are generated exclusively from the available real points using SMOGN and GPR. Model training is then conducted using the union of the selected real samples and the corresponding synthetic data generated from them. No samples from the inference grid are used during real-data selection, synthetic generation, or training.

When SMOGN and GPR generate synthetic x, y coordinates and synthetic received power, they are simply guessing the relationship based on statistics. For this reason, we inject physical laws that tie these samples together, by adding afterwards to synthetic samples the exact mathematical distances from the LEDs. Therefore, the input feature vector includes the receiver coordinates and its distances from the LED transmitters, while the target is the total received power from all the LEDs. To achieve statistically significant results, we average the results of 100 independent training runs.

Final evaluation is performed exclusively on the second staggered grid comprising 5526 real measurements, which remains entirely unseen throughout the training and synthesis process. This strict separation ensures that performance improvement reflects genuine synthetic-to-real generalization without information leakage or spatial overlapping. We repeat this process on different fractions of real data (101, 303, and 606 out of 1818) to quantify the extent to which synthetic data can compensate for sparse training metrics in realistic VLC deployment scenarios. This design allows systematic assessment of how different synthetic data generation strategies support model training as the density of available real measurements decreases.

In this section, synthetic data generation is evaluated using the MLP model only. The MLP was selected as a representative learning-based estimator based on the comprehensive benchmarking and validation results presented in Section 5. Fixing the model architecture allows the effect of synthetic data augmentation under varying levels of real-data sparsity to be isolated, without confounding factors introduced by differences in model capacity or inductive bias. The objective of this analysis is therefore not to re-benchmark learning architectures, but to assess how synthetic data can support training when real measurements are limited.

Following this methodology, the quantitative results for all evaluated augmentation configurations are summarized in Table 10. A baseline performance limit was first established using the full, high-density experimental dataset comprising all 1818 physical measurements from the first staggered grid. Evaluated across the unseen 5526-point inference set, this dense configuration yielded a baseline Mean Absolute Error (MAE) of 0.14 dBm along the diagonal transect and a global MAE of 0.15 dBm (Figure 15a).

Next, the two distinct augmentation methodologies (SMOGN and GPR) were compared using the reduced subset of 606 samples. Feature-space augmentation via SMOGN (adding 172 synthetically generated points) achieved a diagonal MAE of 0.14 dBm but suffered a degradation in global accuracy (MAE = 0.16 dBm) (Figure 15b). On the other hand, increasing the spatial field with synthetic samples of similar volume that is a 13×13 GPR generated grid, maintained both the diagonal (0.14 dBm) and global (0.15 dBm) MAE measurements (Figure 15c).

Based on the better prediction performance of GPR, we proceed to data sparsity analysis. MLP was trained on 50x50 grids generated by GPR based on the three predefined sparsity levels (Figure 15d-f). When the samples are too sparse (101), GPR does not have enough real anchor points to accurately map the power field resulting from the complex multi-LED topology, resulting in a significant deterioration of the global MAE of the MLP to 0.19 dBm (with a diagonal MAE of 0.16 dBm). Increasing the density of the initial set of samples (303 and 606) GPR produces a high-fidelity synthetic grid that significantly improves the prediction, which matches the initial baseline measurements (Diagonal MAE: 0.14 dBm, global MAE: 0.15 dBm).

These results confirm the viability of the proposed methodological pipeline. Real-world data collection requirement in indoor VLC systems can be reduced by approximately 83% (from 1,818 to 303 physical samples) with minimal loss in the predictive accuracy of the neural model used by using GPR for computational augmentation of spatial datasets.

**TABLE 10 Evaluation of augmentation configurations**

| Configuration | Real Samples | Synthetic Method | Diagonal MAE (dBm) | Global MAE (dBm) |
|---|---|---|---|---|
| (a) Baseline | 1,818 | None | 0.14 | 0.15 |
| (b) SMOGN | 606 | SMOGN | 0.14 | 0.16 |
| (c) GPR | 606 | GPR (13×13) | 0.14 | 0.15 |
| (d) Sparse Seed | 101 | GPR (50×50) | 0.16 | 0.19 |
| (e) Mid Seed | 303 | GPR (50×50) | 0.13 | 0.15 |
| (f) Dense Seed | 606 | GPR (50×50) | 0.14 | 0.15 |

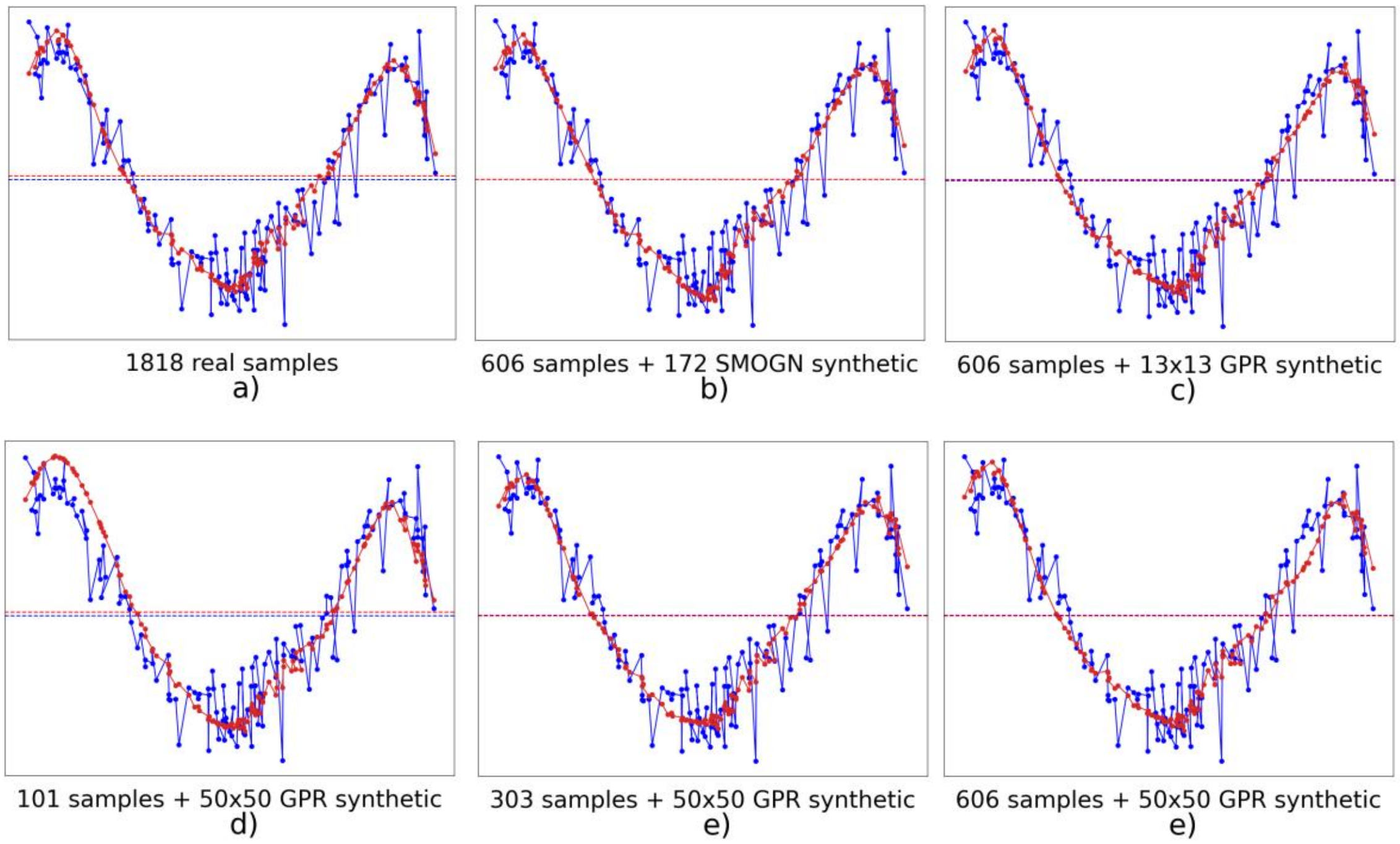

**FIGURE 15** Comparison of ground truth (blue) and MLP-predicted (red) total received power (dBm) along the diagonal. Subplot (a) establishes baseline performance using a dense training set of 1,818 real experimental measurements. Subplots (b) through (f) evaluate the efficacy of various synthetic data augmentation strategies: (b) 606 real samples augmented with 172 SMOGN-generated points in the feature space (c) 606 real samples augmented with a 13×13 GPR-generated spatial grid; and models trained on dense 50×50 GPR-generated grids seeded from (d) 101, (e) 303, and (f) 606 real experimental samples.

## 7 Conclusions

In this work, we propose a data-driven methodology for estimating REM for indoor VLC systems that first evaluates learning-based models using configurable simulated datasets and subsequently benchmarks the selected approach against alternative learning-based and interpolation methods using experimental measurements.

We intentionally use a static indoor environment scenario with fixed transmitter and receiver orientations, without shadowing, blockage, mobility, or higher order multipath effects, throughout this study. This modeling choice is taken to isolate the behavior of data driven REM construction itself and allow for a clean comparison between approaches without confounding influences from factors that exist in complex and dynamic environments. Thus, the chosen setup can serve as a testbed to evaluate accuracy, generalization behavior, robustness, and computational efficiency of VLC REM estimation methods.

A physics-based VLC simulator was built to generate RSS datasets under different room geometries and transmitter configurations, that enabled training and evaluation of Decision Tree ensembles and Multi-Layer Perceptron (MLP) models. MLP architectures consistently achieved lower prediction errors than Decision Trees based approaches and required smaller training datasets, across simulated scenarios. MLP models demonstrated a flexible trade-off between computational cost and accuracy, as they were able to maintain low prediction error and significantly reduce training time by adjusting the training sample size, number of epochs, and batch size. More specifically, small sample sets (4k points) trained for 250 epochs with batch size of 128 provide ~10s training time and extremely fast prediction with very good accuracy, MAE is less than 0.3 dBm for the 3D case in variable length rooms.

Benchmarking on experimental measurements showed that the MLP model achieves prediction accuracy comparable to other high-performing methods, with MAE values around 0.24–0.25 dBm under staggered-grid sampling. Performance degradation was revealed for all methods in spatially blocked cross-validation when extrapolating to unseen regions, with MLP yielding 0.80 ± 0.04 dBm MAE, compared to 0.62 ± 0.02 dBm for CNN-based models and significantly larger errors for interpolation and kernel-based techniques. Runtime analysis demonstrated that MLP provides a favorable balance between accuracy and efficiency, requiring approximately 6.36 s training time, 0.19 s grid-level inference, and ≈31 ms single-point prediction, while physics-based simulation required substantially longer grid-level and single-point computation.

Robustness analysis on sensor jitter and label noise showed gradual degradation of MLP performance, where MAE increases from 0.49 dBm to 1.89 dBm for position perturbations between 0.2 m and 1 m and remained competitive with other neural and kernel-based models. MLP capacity ablation suggests that moderate size (64×256) chosen architecture provides sufficient capacity for VLC REM estimation, as increasing network depth or width does not improve accuracy, and all tested MLP configurations maintained approximately 0.25 dBm MAE.

To address cases with sparse available measurements we investigated synthetic data generation using SMOGN and GPR. Results showed that GPR-based spatial augmentation preserves prediction accuracy with reduced training data. Using only 303 real samples, compared to 1,818 measurements required in the baseline case, GPR maintained 0.14 dBm diagonal MAE and 0.15 dBm global MAE.

Future work will extend the proposed methodology to more realistic indoor VLC scenarios by incorporating dynamic effects such as mobility, receiver orientation variability, shadowing, blockage, and higher-order multipath propagation, as well as exploring adaptive and online learning strategies. Further investigation of advanced synthetic data generation schemes and validation on larger, multiroom experimental datasets will also be required to assess scalability and applicability in operational VLC deployments.

Overall, this study proposes a methodology that first evaluates a data-driven REM approach using simulated data and then benchmarks it against learning-based and interpolation methods using experimental measurements. It demonstrates the practical advantage of data-driven REM approaches, which can achieve comparable or superior performance at a fraction of the computational cost required by a full physics-based simulation. Finally, it is shown that among the evaluated methods, MLP offers the most balanced combination of accuracy, spatial generalization, robustness, and efficiency and support its suitability as a practical basis for VLC REM estimation in controlled indoor environments.

## Acknowledgment

H. S. and C.T.P would like to thank the reviewers because their feedback helped us to extend our work, clarify key concepts and strengthen our arguments. We also thank Dr. Christos Tselios for fruitful discussions on results. Finally, we acknowledge Empeirikion Foundation for funding this work.